\documentclass[letterpaper,12pt,twoside]{article}
\usepackage[T1]{fontenc}
\usepackage[noblocks]{authblk}
\usepackage[letterpaper, top=3cm, bottom=3cm, left=2.5cm, right=2.5cm]{geometry}
\usepackage{natbib}
\usepackage[english]{babel}
\usepackage[T1, OT1]{fontenc}
\DeclareTextSymbolDefault{\dh}{T1}
\usepackage[titletoc,page]{appendix}
\usepackage[utf8]{inputenc}
\usepackage{amsmath}
\usepackage{amsthm}
\usepackage{amsfonts}
\usepackage{fancyhdr}
\usepackage[onehalfspacing]{setspace}
\usepackage{tocbibind}
\usepackage{tocloft} 
\usepackage{titlesec, blindtext}
\usepackage{caption}
\usepackage{subcaption}
\usepackage[normalem]{ulem}  
\usepackage{bm}
\usepackage[final]{showlabels}
\usepackage{nicefrac}
\usepackage[section]{placeins}
\usepackage{multirow}
\usepackage{xcolor}
\usepackage{adjustbox}
\usepackage{tikz}
\usepackage{fancyhdr} 
\usepackage{placeins}
\usepackage[tableposition=top]{caption}
\usepackage{url}
\usetikzlibrary{arrows}
\usetikzlibrary{shapes.geometric}
\definecolor{mygrey}{gray}{0.50}
\newcommand{\indep}{\stackrel{\mathrm{indep}}{\sim}}

\newcommand{\boldW}{\mathbf{W}}
\newcommand{\boldX}{\mathbf{X}}
\newcommand{\boldXta}{\mathbf{X^{\mathrm{ta}}}} 

\newcommand{\boldXpsu}{\mathbf{X^{\mathrm{psu}}}}
\newcommand{\boldXhh}{\mathbf{X^{\mathrm{hh}}}}
\newcommand{\boldXind}{\mathbf{X^{\mathrm{ind}}}}
\newcommand{\boldxind}{\mathbf{x^{\mathrm{ind}}}}
\newcommand{\boldV}{\mathbf{V}}
\newcommand{\boldv}{\mathbf{v}}
\newcommand{\boldx}{\mathbf{x}}

\newcommand{\bolds}{\mathbf{s}}
\newcommand{\boldt}{\mathbf{t}}

\newcommand{\boldbeta}{\bm{\beta}}

\newcommand{\boldalpha}{\bm{\alpha}}
\newcommand{\boldphi}{\bm{\phi}}

\newcommand{\boldxi}{\bm{\xi}}

\newcommand{\Maori}{M$\overline{\mathrm{a}}$ori }

\begin{document}

\title{\bf{Small domain estimation of census coverage: A case study in Bayesian analysis of complex survey data}}
\author{Joane S. Elleouet}
\author{Patrick Graham}
\author{Nikolai Kondratev}
\author{Abby K. Morgan}
\author{Rebecca M. Green}
\affil{Stats NZ, 8 Gilmer Terrace, PO Box 2922, Wellington 6011, New Zealand}
\date{}

\maketitle
\thispagestyle{fancy}

\begin{abstract}
Many countries conduct a full census survey to report official population statistics. As no census survey ever achieves 100\% response rate, a post-enumeration survey (PES) is usually conducted and analysed to assess census coverage and produce official population estimates by geographic area and demographic attributes. Considering the usually small size of PES, direct estimation at the desired level of disaggregation is not feasible. Design-based estimation with sampling weight adjustment is a commonly used method but is difficult to implement when survey non-response patterns cannot be fully documented and population benchmarks are not available. We overcome these limitations with a fully model-based Bayesian approach applied to the New Zealand PES. Although theory for the Bayesian treatment of complex surveys has been described, published applications of individual level Bayesian models for complex survey data remain scarce. We provide such an application through a case study of the 2018 census and PES surveys. We implement a multilevel model that accounts for the complex design of PES. We then illustrate how mixed posterior predictive checking and cross-validation can assist with model building and model selection. Finally, we discuss potential methodological improvements to the model and potential solutions to mitigate dependence between the two surveys.
\end{abstract}

\section{Introduction
 \label{sec:intro}
}

In Aotearoa New Zealand a census is conducted every five years. It is a key input to official population estimates and supports a wide range of social and demographic analyses. Although the census would ideally count all people and their attributes of interest in the country at a given time, it inevitably fails to enumerate the full population. Censuses are expensive undertakings and the performance of the census in enumerating the population is therefore a matter of public interest. Consequently, a post-censal survey (the post-enumeration survey, henceforth PES) is conducted to evaluate the population coverage of the census. As well as providing an evaluation of the census, coverage estimates of the New Zealand census are used to adjust census counts in order to produce population estimates in the form of an estimated resident population (ERP) which is highly disaggregated to geographic and demographic groups. The population estimation system run by Stats NZ (New Zealand' s official statistics agency) therefore requires coverage adjustments at a high level of granularity defined by, at least, combinations of age in single year intervals, sex, ethnicity, 88 local government areas, \Maori descent and country of birth (New Zealand or other). As these variables can potentially form hundreds of thousands of observed domains and the PES sample is made of approximately 30,000 individual records, direct estimation meets crucial limitations and the problem is best viewed as a modelling problem in which the objective is to relate the coverage probability to the covariates of interest. We therefore propose a Bayesian multilevel modelling approach to census coverage estimation. Although the official 2018 population estimates were created using a similar method \citep{PES2018}, the data and models used here differ from those used for the official published census coverage estimates and should not be regarded as official statistics.

Many countries with traditional census collections run a post-census coverage survey. Published applications include \citet{Brownonscoverage,chipperfield2017estimating,Hogan1993USPES1990,Mule2007using}, with coverage estimation methods ranging from adaptations of dual-systems estimation using a variant of the well-known Lincoln-Petersen estimator \citep{Brownonscoverage} to logistic regression of census coverage \citep{Mule2007using,chen2010local} followed by inverse coverage probability weighting of the census file to obtain population estimates. Our methods resemble the latter approach, though we use multilevel logistic models to obtain coverage and population estimates at a high level of granularity. Hierarchical Bayes models have also been proposed for estimation of the coverage of the Canadian census \citep{YouASAhierarchical}. However, these are area-level models, in contrast to the individual level models discussed in this paper. \citet{elliott2000bayesian} developed  a Bayesian  model for  census  coverage estimation that incorporates information on  population sex ratios, in addition to data  from the census  and a census  coverage survey. However, the data structure assumed in that work differs from the one available for the current analysis.

Modelling complex survey data at the individual level requires attending to the impact of the survey design and non-response on inclusion in the data. Whereas the design-based approach to survey inference achieves this through the use of survey weights and variance calculations that respect the survey design, the model-based approach accommodates the impact of survey design and non-response on inclusion in the observed data within the model structure. The latter approach is often accompanied by the application of model-derived estimates to benchmark population data to obtain small domain estimates that account for differences in covariate structure between the sample and the target population, as illustrated by the so-called MRP (Multilevel Regression and Post-stratification) method \citep{gelman1997poststratification,laxandphillips2009,si2017bayesian}. We cannot use population benchmarks to aid coverage estimation from PES because one of the purposes of PES is to adjust the census data to produce new population benchmarks. Nevertheless, the application of highly disaggregated model-derived estimates from PES to the census to produce estimates of the usually resident population has some parallels with the MRP approach to estimation. 

Although the general Bayesian approach to analysis of complex survey data has been well described (\citealp[chapter ~2]{rubin1987multiple}; \citealp{little2003bayesian}; \citealp[chapter ~8]{BDA3}), published applications of individual level Bayesian analyses of complex sample surveys remain relatively rare. Some recent applications, unrelated to census coverage, include small area official statistics \citep{nandram2018bayesian}, political sciences \citep{ghitza2013deep, shirleygelmansurveys2015}, and public health \citep{paige2019design}, the latter using simulations to compare design-based to model-based approaches. Bayesian methods, and particularly multilevel Bayesian models, have more commonly been applied to area-level modelling of complex survey data for small domain estimation. In such applications, summary direct estimates with an associated variance estimate are first computed for each area and/or group of interest. Multilevel Bayesian models are then applied to smooth the summary statistics. In the case of complex survey data the direct estimates and variance estimates computed as the first stage of this procedure are usually design-based estimates. Examples include \citet{ghosh1998generalized,you2006small,molina2014small,chen2014use}. Reviews of the general approach can be found in \citet[pp. 45-47]{pfeffermann2013new} and \citet[chapter 10]{rao2015small}. In this approach, design-based estimation is used to deal with the analytical complications of complex sample surveys, freeing the multilevel Bayesian modelling from the requirement to explicitly deal with the survey design.

Application of the area-level approach is problematic in our context where the number of covariate combinations (or domains) exceeds the number of records in the survey dataset, so that forming the initial set of domain-level summary statistics is not even possible. Even applying the area-level approach to an aggregated version of the cross-classification of covariates for which estimates are ultimately required, such that each covariate combination in the aggregated cross-classification occurs in PES, would be difficult unless the degree of coarsening is substantial. In sparse data situations with a binary outcome, conventional design-based variance estimates of proportions can often be zero and this makes subsequent modelling difficult. Consequently, framing the problem as estimation from a model fitted at the level of individual records and from which predictions can then be made seems a logical way forward. However, accounting for a complex survey design complicates the model so that the model fitted to the data is more complex than required for prediction. We illustrate how the model of interest can be, implicitly, recovered from the fitted model by integrating out parameters associated with the survey design but not relevant to the predictions. This paper illustrates the potential of Bayesian modelling of complex survey data for challenging small domain estimation problems.

To describe our approach, we first describe the PES design in Section \ref{sec:PESdescription}. We then present our modelling strategy in Section \ref{sec:pesest}, including model-checking and evaluation. In Section \ref{sec:results}, we show results of the model checking procedure. We also include summaries of standardised coverage estimates, by area and by age and ethnic group.
The standardisation is achieved by applying the modelled coverage estimates for each group to a common reference population. The reference population used for this estimation is the population estimated by adjusting the census file for under-coverage using the disaggregated coverage estimates obtained from the model. Uncertainty in the estimation of the reference population is automatically incorporated in the posterior distribution for the standardised estimates. Section \ref{sec:discuss} concludes the paper with some discussion of the modelling issues and suggestions for further development.

\section{PES and Census Data}
\label{sec:PESdescription}

The official 2018 census dataset comprises census respondents as well as records obtained from administrative sources \citep{census2018overview}. It is subject to under-coverage (eligible residents missed by the census) and over-coverage (non-eligible individuals mistakenly counted, such as births after the census date and residents temporarily overseas at the time of census). In the 2013 census, over-coverage was approximately 0.7 \%, in contrast to an under-coverage measure of approximately 3.1\% \citep{PES2013}. The official ERP is corrected for both types of errors estimated on the full census file \citep{PES2018}. Estimates presented here differ from previously published estimates of the population coverage of the official census file and should not be regarded as official statistics. The main differences with the methodology used for official statistics is that we focus on under-coverage probability estimation and we perform the estimation on the respondent subset of the census file, which excludes administrative enumerations. However, the estimation challenge we describe is similar to the one faced in constructing the official 2018 ERP. Estimates for under-coverage probabilities hold without having to make assumptions about levels of over-coverage. \citet{PES2018} addresses over-coverage estimation in a very similar manner to under-coverage, and we refer the reader to this publication for more details on over-coverage estimation.  
 
The 2018 PES used an area-based, stratified two-stage design. For sampling purposes New Zealand was divided into 23,174 small geographic areas (Primary Sampling Units, PSUs) that were grouped into 101 strata, based on a combination of broad geographic region, major urban status, census delivery mode (whether an access code for the online census form was mailed out or a hard copy census form was delivered) and a measure of deprivation. The PSUs were selected using probability proportional to size (PPS), where the size measure was based on historical estimates and included an adjustment for ethnic group proportions. Sampling fractions varied by strata, with urban strata sampled more intensively than non-urban strata, for fieldwork efficiency reasons. PES operated in all strata and a total of 1,365 PSUs were selected for the PES sample. In most PSUs, 11 dwellings were sampled within each PSU using Stats NZ's standard approach in which dwellings within a PSU are grouped into panels of size 11, and one panel is randomly selected. This resulted in a sample of 15,213 households within 15,015 dwellings in the 1,365 selected PSUs. Dwellings refer to the building in which people live, whereas people residing together and sharing facilities within a dwelling constitute a household, and there can be multiple households per dwelling. All usual residents at selected households were eligible for inclusion in the sample. Henceforth, we refer to households and use the terms household effects and household variables when referring to both household and dwelling characteristics. Within the 15 213 visited households, 37,548 people were interviewed. After filtering for refusal, incomplete responses and ineligibility, the final sample included 12,459 households and 31,600 respondents with responses of sufficient quality to be linked to the census and included in the estimation.

The PES sample was linked to the census file using a conservative probabilistic linking methodology, followed by clerical checking of all non-linked records and a sample of linked records. Details are described in \citet{PES2018}. Of all eligible PES person records, 30,397 were linked to a census record (1,300 through manual linking), and the remaining 1,203 PES respondents were not linked to any census record. PES respondents linked to a record in the census respondent file were considered covered by census, whereas unlinked PES records constitute instances of under-coverage.

\section{Census coverage estimation under a Bayesian modelling framework}
 \label{sec:pesest}

\subsection{The Bayesian approach}

Fully model-based analysis of complex survey  data usually requires multilevel models in order to account for the survey design. Such modelling  fits neatly  into a Bayesian framework. The Bayesian approach to inference permits coherent assessment of uncertainty  for all model  parameters and provides  a flexible framework  for propagation of parameter uncertainty to quantities derived from the model. We exploit this flexibility  to obtain posterior  distributions for highly disaggregated  coverage probabilities (see Section \ref{subsec:predict}) and for useful summaries of these probabilities (see Section \ref{res:stand}).

We generally  specify prior distributions to be only weakly informative, in the sense of being open-minded as to  the range of parameter values, while guaranteeing that inherent  range constraints are respected (e.g. positive variances)  and discouraging, but not disallowing, extreme values \citep{gelman2008weakly}. 

In our application, we obtain a Monte Carlo approximation to the joint  posterior distribution for all model  parameters, by generating a sample from  the posterior using Markov Chain Monte Carlo (MCMC) methods. Specifically, the sample is obtained using the program Stan \citep{2020stan} through the R interface \citep{2020rstan,rcite}. Stan implements Hamiltonian Monte Carlo, a popular type of MCMC algorithm known to reduce the correlation between successive sampled values and, therefore, efficiently converging to the posterior distribution.

\subsection{General assumptions}

A critical assumption of Bayesian analysis of survey data is ignorability (\citealp[chapter ~2]{rubin1987multiple}; \citealp{little2003bayesian}; \citealp[chapter ~8]{BDA3}), which in the case of PES, requires conditional independence of inclusion in PES and inclusion in census, given the model covariates and \emph{a priori} independence of the parameters of the models for inclusion in PES and in census. 
The former assumption is similar to the often invoked ``independence" assumption of dual systems population estimation \citep{sekar1949method,Brownonscoverage}. When ignorability holds, inference for inclusion in census (that is, census coverage) can proceed without specifying and fitting the model for inclusion in PES. 
In order to justify the assumption of ignorability, it is usually necessary to include the survey design features in the model, along with other covariates associated with non-response. We follow this approach in developing the model for census coverage. 
The nested geographical clustering of the sample design naturally lends itself to multilevel modelling, and, fortunately, in our case, there is overlap between variables of substantive interest and those predictive of non-response. 
We discuss the ignorability assumptions for our analysis in more detail in section C of the Supplementary Material, which tailors the general approach to Bayesian analysis of  complex surveys given in \citet[chapter 8]{BDA3} to the specific case of PES.
 
 As well as conditional independence of inclusion in PES and census, we make the other standard assumptions of dual systems population estimation. We assume no errors in the linkage of PES to census, and we assume the target population is closed over the operating periods of census and PES.

\subsection{General under-coverage model} \label{subsec:modelspec}

We let $\boldX$ denote demographic covariates and $\boldx$ a particular covariate combination.
We use the notation $\mathrm{TA}$ to denote geographic area, and let $t \in \{1,\ldots, 88\}$ indicate a particular TA. To simplify notation in this section we let $\boldV = (\boldX,\mathrm{TA}),$ so $\boldv=(\boldx,t)$ refers to a particular covariate combination $\boldx$ in TA $t$. The sample space for $\boldV$ is the space of all covariate-TA combinations, denoted $\cal{V}.$

Introducing the indicators $C$ and $Q$ for inclusion in the census and in the target population respectively, we define the under-coverage probability as 
\begin{equation*}
p_{under}(\boldv,\boldxi) = \Pr(C=0 |Q =1, \boldV=\boldv,\boldxi),
\label{eq:punder}
\end{equation*}
where $\boldxi$ is the parameter vector of the under-coverage model.

The purpose of the model presented here is to estimate $p_{under}(\boldV,\boldxi)$. A coverage-adjusted population estimate based on the census can subsequently be obtained by weighting each census record by the inverse of the under-coverage probability:

\begin{equation}
w_{i} = \frac{1}{(1 - p_{under}(\boldv_{i},\boldxi))},
\label{eq:car}
\end{equation}
where the subscript refers to the $i^{th}$ census respondent. Using a Bayesian approach enables this adjustment to be applied to each census record for each draw from the posterior distribution for $p_{under}(\boldv_{i},\boldxi)$ in a Monte Carlo procedure which produces as many simulations of the ERP as needed to obtain precise uncertainty measures (e.g. approximate credible intervals). More details on the Monte Carlo methodology of the ERP production can be found in \citet{bryant2016measuring} and \citet{ERP2018}.

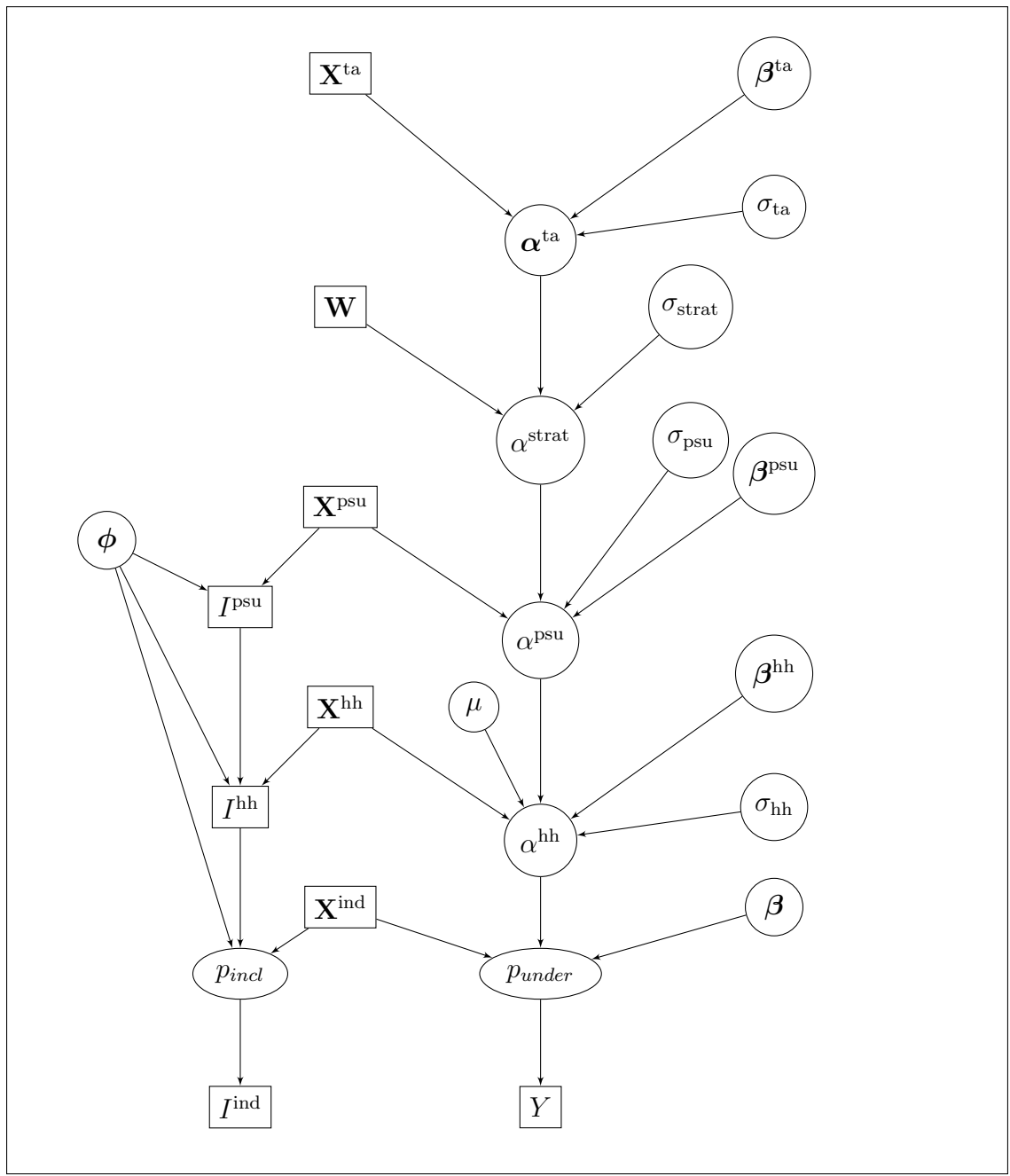
\begin{figure}
	\centering
	\begin{tikzpicture}
	\tikzset{prm/.style = {shape=circle,draw,minimum size=1.5em}}
	\tikzset{edge/.style = {->,> = latex'}}
	\tikzset{undirect/.style = {-,> = latex'}}
	\tikzset{data/.style = {shape=rectangle,draw,minimum size=1.5em}}
	\tikzset{det/.style = {shape=ellipse,draw,minimum size=1.5em}}
	\node[data] (Xta) at (-3,8.5) {$\boldX^{\mathrm{ta}}$};
	\node[prm]  (betata) at (3.5,8.5) {$\boldbeta^{\mathrm{ta}}$};
	\node[prm]  (sigmata) at (3.5,6.5) {$\sigma_{\mathrm{ta}}$};
	\node[prm]  (alphata) at (0, 6) {$\boldalpha^{\mathrm{ta}}$};	
	\node[data]  (W)   at (-3,5) {$\boldW$};
	\node[prm]   (alphastrat) at (0,3) {$\alpha^{\mathrm{strat}}$};
	\node[prm]   (sigmastrat) at (2.25,5) {$\sigma_{\mathrm{strat}}$};
	\node[prm]  (alphapsu) at (0,0) {$\alpha^{\mathrm{psu}}$};
	\node[data]  (Xpsu)   at (-3,2) {$\boldX^{\mathrm{psu}}$};
	\node[data]  (Ipsu)  at (-4.5,0.5) {$I^{\mathrm{psu}}$};
	\node[prm]  (phi)   at (-6.5,1.5) {$\boldphi$};
	\node[prm] (betapsu) at (3.5,2.5) {$\boldbeta^{\mathrm{psu}}$};
	\node[prm]  (sigmapsu) at (2.25,3)  {$\sigma_{\mathrm{psu}}$};
	\node[prm]  (alphahh) at (0,-3) {$\alpha^{\mathrm{hh}}$};
	\node[data]  (Xhh) at (-3,-1) {$\boldX^{\mathrm{hh}}$};
	\node[data]  (Ihh) at (-4.5,-2.5) {$I^{\mathrm{hh}}$};
	\node[prm]  (betahh) at (3.5,-0.5) {$\boldbeta^{\mathrm{hh}}$};
	\node[prm]  (sigmahh) at (3.5,-2.5) {$\sigma_{\mathrm{hh}}$};
	\node[prm] (mu)  at (-1.0,-1) {$\mu$};
	\node[det] (punder) at (0,-5) {$p_{under}$};
	\node[data]  (Y) at (0,-7)  {$Y$};
	\node[data] (X) at (-3.0,-4)  {$\boldX^{\mathrm{ind}}$};
	\node[prm] (beta) at (3.5,-4) {$\boldbeta$};
	\node[det] (pincl) at (-4.5,-5) {$p_{incl}$};
	\node[data] (Iindiv) at (-4.5,-7) {$I^{\mathrm{ind}}$};
	\draw[edge] (Xta) to (alphata);
	\draw[edge] (betata) to (alphata);
	\draw[edge] (sigmata) to (alphata);
	\draw[edge] (W) to (alphastrat);
	\draw[edge] (alphata) to (alphastrat);
	\draw[edge] (sigmastrat) to (alphastrat);
	\draw[edge] (alphastrat) to (alphapsu);
	\draw[edge] (Xpsu) to (alphapsu);
	\draw[edge] (betapsu) to (alphapsu);
	\draw[edge] (alphapsu) to (alphahh);
	\draw[edge] (Xhh) to (alphahh);
	\draw[edge] (betahh) to (alphahh);
	\draw[edge] (sigmahh) to (alphahh);
	\draw[edge] (mu) to (alphahh);
	\draw[edge] (alphahh) to (punder);
	\draw[edge] (X) to (punder);
	\draw[edge] (beta) to (punder);
	\draw[edge] (punder) to (Y);
	\draw[edge] (X) to (pincl);
	\draw[edge] (Xhh) to (Ihh);
	\draw[edge] (Ihh) to (pincl);
	\draw[edge] (Xpsu) to (Ipsu);
	\draw[edge] (Ipsu) to (Ihh);
	\draw[edge] (phi) to (Ipsu);
	\draw[edge] (phi) to (Ihh);
	\draw[edge] (phi) to (pincl);
	\draw[edge] (pincl) to (Iindiv);
	\draw[edge] (sigmapsu) to (alphapsu);
	\draw (-8,9.5) rectangle (7.0, -8);
	\end{tikzpicture}
	\caption{Directed Acyclic Graph (DAG) representing the structure of the model for census coverage ($Y$). Rectangles represent observables and circles represent model parameters. $p_{incl}$ and $p_{under}$ are probabilities obtained as deterministic functions of observables and parameters. $I^{\mathrm{psu}}$, $I^{\mathrm{hh}}$, $I^{\mathrm{ind}}$ are indicators for PSU, household and individual inclusion in PES. The inclusion model depends on the parameter vector, $\boldphi$. The inclusion indicators may depend on the model covariates but, because they are assumed conditionally independent of census coverage and parameters of the census coverage model, given the covariates, and because the inclusion and coverage model parameters are assumed to be \emph{a priori} independent, inclusion in the PES is ignorable. Under these assumptions, modelling of census coverage using PES data can proceed without modelling inclusion in PES.}
	\label{fig:dag}
\end{figure}

We let $N_{h}^{\mathrm{ind}}$, denote the number of usual residents within household $h$, $N_p^{\mathrm{hh}}$, the number of households in PSU $p$, $N_s^{\mathrm{psu}}$, the number of PSUs in stratum $s$, $N_t^{\mathrm{strat}}$, the number of strata intersecting TA $t$, $N_{\mathrm{tot}}^{\mathrm{strat}}$ the total number of strata and $N^{\mathrm{ta}}$, the total number of TAs. After the linking procedure between PES and census, each record $j$ in household $h$ in the PES dataset receives an under-coverage indicator $Y_{hj}$ which states whether the record is present in the census file ($Y=0$) or absent from it ($Y=1$). Each record is also characterised by a set of demographic covariates $\boldX_{hj}^{\mathrm{ind}}$, geographic variables related to the survey design, and local government area, $\mathrm{TA}$. We present the model for census under-coverage in two ways: with a directed acyclic graph (DAG) (Figure \ref{fig:dag}), and with the following equations, followed by a description.

\begin{align}
  \left[Y_{{hj}} | \boldX_{hj}^{\mathrm{ind}},p_{under_{hj}} \right] &\indep \textrm{Bernoulli} \left(p_{under_{hj}} \right); j=1,...,N_h^{\mathrm{ind}} ; h=1,..., N_{\mathrm{psu}[h]}^{\mathrm{hh}}, \label{eq:bern}\\
  \mathrm{logit}(p_{under_{hj}}) &= \alpha_{h}^{\mathrm{hh}} + \boldX_{hj}^{\mathrm{ind}'}\beta ; \, j=1,...,N_h^{\mathrm{ind}} ; h=1,..., N_{\mathrm{psu}[h]}^{\mathrm{hh}}, \label{eq:logit} \\
  \begin{split}
  \left[ \alpha_h^{\mathrm{hh}} | \boldX_{h}^{\mathrm{hh}},\alpha_{\mathrm{psu}[h]}^{\mathrm{psu}}, \boldbeta^{\mathrm{hh}} , \sigma_{\mathrm{hh}}^2 \right] & \indep \mathcal{N} \left(\mu + \alpha_{\mathrm{psu}[h]}^{\mathrm{psu}} + %
  \boldX_{h}^{\mathrm{hh}'}\boldbeta^{\mathrm{hh}}, \sigma_{\mathrm{hh}}^2 \right) ; h:\mathrm{psu}[h]=p; \\%
  & ~\phantom{\mathcal{N} \left(\mu + \alpha_{\mathrm{psu}[h]}^{\mathrm{psu}} + %
  \boldX_{h}^{\mathrm{hh}'}\boldbeta^{\mathrm{hh}}, \sigma_{\mathrm{hh}}^2 \right) ;abcd} p=1,..., N_s^{\mathrm{psu}}, 
  \end{split}
  \label{eq:hh} \\
  \begin{split}
  \left[ \alpha_{p}^{\mathrm{psu}} | \boldX_{p}^{\mathrm{psu}},\alpha_{\mathrm{strat}[p]}^{\mathrm{strat}}, \boldbeta^{\mathrm{psu}} , \sigma_{\mathrm{psu}}^2 \right] &\indep \mathcal{N} \left(\alpha_{\mathrm{strat}[p]}^{\mathrm{strat}} + \boldX_{p}^{\mathrm{psu}'}\boldbeta^{\mathrm{psu}}, \sigma_{\mathrm{psu}}^2 \right) ; p: \mathrm{strat}[p]=s ; \\ %
 & ~ \phantom{\indep \mathcal{N} \left(\alpha_{\mathrm{strat}[p]}^{\mathrm{strat}} + \boldX_{p}^{\mathrm{psu}'}\boldbeta^{\mathrm{psu}}, \sigma_{\mathrm{psu}}^2 \right) ; } s=1,..., N_t^{\mathrm{strat}}, 
 \end{split}
 \label{eq:psu} \\
  \left[\alpha_s^{\mathrm{strat}} | \boldW,\boldalpha^{\mathrm{ta}}, \sigma_{\mathrm{strat}}^2\right] &\indep \mathcal{N}\left(\boldW_{s} \boldalpha^{\mathrm{ta}}, \sigma_{\mathrm{strat}}^2\right) , s=1, \ldots, N_{\mathrm{tot}}^{\mathrm{strat}}, \label{eq:strat} \\
  \left[\alpha_{t}^{\mathrm{ta}}|\boldX_{t}^{\mathrm{ta}},\boldbeta^{\mathrm{ta}},\sigma_{\mathrm{ta}}^2 \right] &\indep \textrm{$t_3$} \left(\boldX_{t}^{\mathrm{ta}'} \bm{\beta}^{\mathrm{ta}}, \sigma_{\mathrm{ta}}^2\right), t=1,...,N^{\mathrm{ta}}, \label{eq:ta}
\end{align}
where the notation $\mathrm{psu}[h],$ and $\mathrm{strat}[p]$ refer, respectively, to the PSU of the $h^{th}$ household and the stratum of the $p^{th}$ PSU. Note $\boldalpha^{ta}=(\alpha_{1}^{ta},\ldots,\alpha_{N^{\mathrm{ta}}}^{\mathrm{ta}} )'$ is a $N^{\mathrm{ta}} \times 1$ vector of TA effects. The notation $t_3$ in equation (\ref{eq:ta}) corresponds to a Student-$t$ distribution with three degrees of freedom.

We model the under-coverage indicator $Y_{hj}$ for individual $j$ in household $h$ using a Bernoulli distribution with probability $p_{under_{hj}}$ (\ref{eq:bern}). 
A logistic regression is specified for $p_{under_{hj}}$ with individual covariates $\boldX_{hj}^{\mathrm{ind}}$ and household-specific varying intercept $\alpha_{h}^{\mathrm{hh}}$ (\ref{eq:logit}). 
Equations (\ref{eq:hh})-(\ref{eq:strat}) show how the varying intercept $\alpha_{h}^{\mathrm{hh}}$ contains an overall average $\mu$ and all levels of the hierarchy reflecting the PES sampling design: it is modelled as a normal distribution, and the mean of this distribution is the result of a regression with a PSU-level effect and household covariate effects (\ref{eq:hh}). 
These household covariates are a ``hard-to-find" binary variable (HTF) which accounts for the variation in dwelling enumeration success between areas (details in section D of the Supplementary Material), and potential individual demographic variables summarised at the household level. 
The PSU-level effect $\alpha_{p}^{\mathrm{psu}}$ is itself a varying effect that we model with a normal distribution, and the mean of this distribution is the result of a regression with a stratum-level effect term and PSU covariate effects (\ref{eq:psu}). 
These covariates are the PSU sampling variables used in the PES sampling design (see Table \ref{table:covariates}). The stratum-level effect $\alpha_s^{\mathrm{strat}}$ is a varying effect modelled with a normal distribution whose mean is a weighted mean of TA-level effects from TAs present in the stratum (\ref{eq:strat}). 
We add a TA level to the model, as this is the geographic resolution required for the publication of the ERP. 

As each of the 101 strata generally spans several TAs, the relationship of strata to TAs is described by an occurrence matrix $\boldW$ where each row corresponds to a stratum and each column corresponds to a TA. 
Each matrix cell $\boldW(s,t)$ therefore represents the proportion of TA $\boldt$ included in stratum $\bolds$. These cell proportions are estimated based on individual counts within small geographical units in the augmented census file, which includes administrative records in addition to census responses \citep{census2018overview}. 
We let $\boldW_{s}$ denote the $s^{th}$ row of $\boldW$. Finally the TA effects $\alpha_{t}^{\mathrm{ta}}$ are modelled through covariates $\boldXta$, which correspond to four socio-economic predictors of TA effects that are calculated from NZ Deprivation indices \citep{NZDEP2018}. We choose a $t_{3}$ distribution at the TALB level because it has more mass in its tails than the normal distribution, which helps avoid over-shrinkage at higher levels of the hierarchical model. 
A detailed description of the individual covariates included in $\boldX^{\mathrm{ind}}$ as well as higher level covariates $\boldXhh$, $\boldXpsu$ and $\boldX^{\mathrm{ta}}$ is given in Table \ref{table:covariates}.

In practice, incorporating the group covariates $\boldXpsu$ and $\boldXhh$ at the individual level of the model (\ref{eq:logit}) by allocating all individuals in a group (household or PSU) the covariate values for that group gives an equivalent formulation to (\ref{eq:bern} - \ref{eq:ta}) and makes subsequent predictions easier to compute. We let $\boldX = (\boldXind', \boldXhh', \boldXpsu')'$. A demonstration of the equivalence of the two approaches is detailed in section B of the Supplementary Material.

The choice of individual covariates used in (\ref{eq:logit}) is largely guided by New-Zealand post-enumeration surveys from previous years (Table \ref{table:covariates}). For instance, age, ethnic group and sex are known to affect census inclusion in distinct ways, so we include these variables and interactions in all models we examine. The four ethnic indicators are \Maori, Pacific, Asian and Other. They are mutually non-exclusive, allowing individuals to belong to multiple ethnic groups. We added interaction terms between ethnicities representing two common profiles of people with multiple ethnicities: \Maori-Other, and \Maori-Pacific. For individual-level variables that are available but whose effect on coverage is less obvious, and for more subtle interactions between covariates, we compute several models differing in their covariates and interactions. Careful examination of resulting parameter posterior distributions and predictions as well as out-of-sample deviance calculations are the used to guide model selection.

All individual covariates except age are binary variables. The challenge with modelling age is that it is inherently an ordered categorical variable with potentially more than 100 categories. Treating this variable as such creates the challenge of estimating a large number of parameters, and dividing the sample into excessively small categories. One solution is to create broader categories such as 5-year age groups, but this solution does not reflect the continuous character of age and its effect on census coverage. It also introduces the additional issue of subjectively selecting categories, and creates breaks among contiguous years that may share extreme values. Another solution, implemented here, is to apply a spline transformation to the original variable. We model age using 10 quadratic splines defined by eight internal breakpoints (see Table \ref{table:covariates}). Figure S1 illustrates the transformation by showing the spline values for each age present in the census. One can see that at any given age, a maximum of three splines contribute to describing the underlying age. This stems from our choice of quadratic polynomials rather than higher-degree polynomials, in order to limit the smoothing of patterns that would result from highly overlapping spline curves.

\begin{table}
  \centering
  \begin{adjustbox}{max width=\textwidth}
  \begin{tabular}{p{4cm} p{1cm} p{9cm} p{1cm}}
    \hline
    \textbf{Variable} & \textbf{Coding} & \textbf{Description} & \textbf{n. param} \\ \hline
    \multicolumn{2}{l}{\textbf{individual covariates}} & & \\
    sex & binary & 0=male, 1=female & 1 \\
    age & 10 splines & quadratic age splines with knots at ages 10, 20, 30, 40, 51, 61, 71, and 81 & 10 \\
    M\={a}ori & binary & M\={a}ori ethnicity indicator & 1 \\
    Pacif & binary & Pacific ethnicity indicator & 1 \\
    Asian & binary & Asian ethnicity indicator & 1 \\
    Other & binary & indicator for "other" ethnicities & 1 \\
    NZ born & binary & 0=born abroad, 1=born in New Zealand \\
    M\={a}ori descent & binary & 0=non-M\={a}ori descent, 1=M\={a}ori descent & 1 \\
    & & & \\
    \multicolumn{2}{l}{\textbf{individual covariate interactions}} & & \\
    M\={a}ori * Other & binary & & 1 \\
    M\={a}ori * Pacif & binary & & 1 \\
    sex * age & binary & sex and all 10 age splines & 10 \\
    Asian * NZ born & binary & & 1 \\
    ethnicity * age & binary & 5 first age splines with each ethnicity and with M\={a}ori * Other (3-way) & 25 \\
    & & & \\
    \multicolumn{2}{l}{\textbf{household covariates} $\boldXhh$ \textcolor{mygrey}{(model 2 only)}} & & \\
    \textcolor{mygrey}{M\={a}ori} & \textcolor{mygrey}{binary} & \textcolor{mygrey}{presence of M\={a}ori} & \textcolor{mygrey}{1} \\
    \textcolor{mygrey}{Pacif} & \textcolor{mygrey}{binary} & \textcolor{mygrey}{presence of Pacific} & \textcolor{mygrey}{1} \\
    \textcolor{mygrey}{Asian} & \textcolor{mygrey}{binary} & \textcolor{mygrey}{presence of Asian} & \textcolor{mygrey}{1} \\
    \textcolor{mygrey}{Other} & \textcolor{mygrey}{binary} & \textcolor{mygrey}{presence of Other} & \textcolor{mygrey}{1} \\
    \textcolor{mygrey}{Female} & \textcolor{mygrey}{binary} & \textcolor{mygrey}{presence of females} & \textcolor{mygrey}{1} \\
    \textcolor{mygrey}{M\={a}ori descent} & \textcolor{mygrey}{binary} & \textcolor{mygrey}{presence of people of M\={a}ori descent} & \textcolor{mygrey}{1} \\
    \textcolor{mygrey}{NZ born} & \textcolor{mygrey}{binary} & \textcolor{mygrey}{presence of people born in New Zealand} & \textcolor{mygrey}{1} \\
    HTF & binary & hard-to-enumerate area & 1 \\
    & & & \\
    \multicolumn{2}{l}{\textbf{household covariate interactions} \textcolor{mygrey}{(model 2 only)}} & & \\
    \textcolor{mygrey}{between ethnicity indicators} & \textcolor{mygrey}{binary} & & \textcolor{mygrey}{6} \\
    \textcolor{mygrey}{ethnicity * female} & \textcolor{mygrey}{binary} & & \textcolor{mygrey}{1} \\
    \textcolor{mygrey}{ethnicity * NZ born} & \textcolor{mygrey}{binary} & & \textcolor{mygrey}{1} \\
    & & & \\
    \multicolumn{2}{l}{\textbf{PSU covariates} $\boldXpsu$} & & \\
    Pacif prop & continuous & proportion of Pacific adults & 1 \\
    PSU size & categorical & S($<$ 50 dwellings)/M(50-100)/L($>$100) & 2 \\
    & & & \\
    \multicolumn{2}{l}{\textbf{TA covariates} $\boldXta$} & & \\
    communication & continuous & Prop. of people with no access to internet at home & 1 \\
    income & continuous & Prop. of people living in households with income below the poverty threshold & 1 \\
    qualification & continuous & Prop. of people aged 18-64 without any qualifications & 1 \\
    internet response & continuous & proportion of online census responses & 1 \\ \hline
  \end{tabular}
  \end{adjustbox}
  \caption{Covariates used in the coverage model. “n. param” shows the number of parameters estimated for each of the covariates and covariate interactions. Covariates only used in the second model are depicted in grey.}
  \label{table:covariates}
\end{table}

We select $Cauchy\left(0,2.5\right)$ as a prior for $\mu$, which is a standard prior recommended in \citet{gelman2008weakly}. Covariate effect parameters $\bm{\beta},$ $\bm{\beta}^{\mathrm{hh}},$ $\bm{\beta}^{\mathrm{psu}}$ and $\bm{\beta}^{\mathrm{ta}}$ are drawn from independent $\mathcal{N}\left(0,1\right)$ distributions. This is not unduly restrictive yet places low prior probability on extreme values. As a reference point, after converting to the odds ratio scale, a $\mathcal{N}\left(0,1\right)$ prior for a logistic regression parameter corresponds to the $95\%$ prior interval $\exp(\pm 1.96),$ implying a $\exp(3.92) \approx 50$ - fold range of prior variation for the effect in question. Group-level variances $\sigma_{\mathrm{ta}}^2$, $\sigma_{\mathrm{strat}}^2$, $\sigma_{\mathrm{psu}}^2$, and $\sigma_{\mathrm{hh}}^2$ are drawn from independent $Cauchy^+\left(0,2.5\right)$ distributions, where $Cauchy^+$ refers to the Cauchy distribution truncated to positive values. 

We run three HMC chains for each model, with the first half used as warm-up. We set the target average proposal acceptance probability to 0.9 and let all other algorithm parameters be set at their default value. For each model, we determine chain length experimentally by increasing it until convergence is reached. We ensure convergence by using the potential scale reduction factor, $\hat{R}$ \citep[pp 284-285]{BDA3} and by visually assessing chain profiles. We also monitor the effective Monte Carlo sample size to ensure appropriate post-convergence Monte Carlo sample size \citep[pp 286-287]{BDA3}.

We explore potential models in two stages. We first focus on individual covariates and their interactions. Group-level covariates at the household, PSU and TA level as well as the stratum level are present in the varying intercept to account for the PES sampling design. The basic model therefore involves all individual covariates as well as the group-level covariates pertaining to the sample design. 

However, results associated with this approach (see section \ref{sec:results}) suggest that individual covariates cannot fully account for variation in census coverage. As the census interview process is dwelling-based, households are an important component of the survey design and this is accounted for in the model through the first level of the varying intercept, $\alpha^{\mathrm{hh}}$. It is likely that census response is partly driven by household-level characteristics that are unobserved. It is also possible that an individual's response or non-response is influenced by another individual in the household. For instance, it is reasonable to suggest that children's response to census is dependent on the parents or caregivers they live with. In such cases, we expect non-response of the former to depend on non-response of the latter, therefore bringing non-response at the household level. This is inconsistent with the model structure, which implicitly assumes that the household-level intercept and individual predictors are independent. To allow for correlation between household and individual characteristics, we follow the solution described in \citet[pp 506-507]{gelman2006data}: we create versions of the individual covariates aggregated at the household level. Therefore, in a second stage, we experiment with the creation of many household-level covariates calculated from all individual covariates. The new covariates are included in $\boldX_{h}^{\mathrm{hh}}$ in equation (\ref{eq:hh}), and described in Table \ref{table:covariates}. The outcomes from including these additional household covariates in the model are addressed in sections \ref{sec:results} and \ref{sec:discuss}.

\subsection{Predicting under-coverage probabilities of census records} 
\label{subsec:predict}
To produce the ERP, coverage probabilities are required for each combination of covariates occurring in the census file. Household- and PSU-level covariates are included in the ERP production but not individual household or PSU effects. The geographic level for application of coverage probabilities is the TA level. While other choices could have been made, these settings provided a compromise between computational tractability and granularity of estimation. Below we describe how the model can be used to generate coverage probabilities at the desired level of demographic and geographic detail.

After fitting the multilevel logistic model, 1000 samples are extracted from the posterior distribution. Each of the 1000 draws from the posterior can be used to predict the under-coverage probability associated with each combination of covariate values that exist in the census. Parameters related to the sampling design (household, PSU, and stratum effects) are integrated to obtain a posterior prediction for each covariate-TA combination. For each combination of TA and individual, household and PSU level covariates, $\boldv=(\boldx',t)'$ where $\boldx=(\boldx^{\mathrm{ind}\prime},\boldx^{\mathrm{hh}\prime},\boldx^{\mathrm{psu}\prime})',$  and for each draw from the posterior of $\boldxi$, we require
\begin{align}
 & p_{\mathrm{under}}( \boldv,  \boldxi)  = \Pr(Y=1 |\boldV=\boldv,\boldxi) \nonumber \\
    = & \sum\limits_{s=1}^{N_{t}^{\mathrm{strat}}} \Pr(Y=1 \vert \boldX=\boldx, \mathrm{strat} = s, \boldxi) \Pr(\mathrm{strat}=s \vert \mathrm{TA}=t) \label{eq:prediction1} \\
   \begin{split}
 = &  \sum\limits_{s=1}^{N_t^{\mathrm{strat}}} \Bigg (\int \Big ( \Pr(Y=1 \vert \alpha^{\mathrm{hh}}, \boldXind=\boldx^{\mathrm{ind}}, \mathrm{strat}=s, \boldxi)   \,\, \times   \\ 
  & ~  \phantom{ \sum\limits_{s=1}^{N_t^{\mathrm{strat}}} }
    p(\alpha^{\mathrm{hh}} \vert \boldX^{\mathrm{hh}}=\boldx^{\mathrm{hh}},\boldX^{\mathrm{psu}}=\boldx^{\mathrm{psu}}, \mathrm{strat}=s, \boldxi) 
    \Big ) \, \mathrm{d}\alpha^{\mathrm{hh}} \times \Pr(\mathrm{strat}=s \vert \mathrm{TA}=t)  \Bigg )   
   \end{split}
   \label{eq:prediction2}    \\
  = &\sum\limits_{s=1}^{N_t^{\mathrm{strat}}} \Bigg (\int \mathrm{expit}(\alpha^{\mathrm{hh}} + \boldx^{\mathrm{ind}\prime} \boldbeta) \mathcal{N}(\alpha^{\mathrm{hh}} \vert \mu + \boldx^{\mathrm{hh}\prime}\boldbeta^{\mathrm{hh}} +
   \boldx^{\mathrm{psu}\prime}\beta^{\mathrm{psu}} + \alpha_s^{\mathrm{strat}}, \sigma_{\mathrm{hh}}^2 + \sigma_{\mathrm{psu}}^2) \, \mathrm{d}\alpha^{\mathrm{hh}} \nonumber \\ 
   & \qquad \times \Pr(\mathrm{strat}=s \vert \mathrm{TA}=t) \Bigg ), \label{eq:19}
\end{align}
where expit() is the inverse logit function and $\mathcal{N}(.|\mu,\sigma^2)$ is the normal density function with mean $\mu$ and variance $\sigma^2$. Writing $\boldX = \boldx$ instead of $\boldV = \boldv$ in the first component of  equation (\ref{eq:prediction1}) follows from the assumptions of the model given by (\ref{eq:bern} - \ref{eq:ta}). TAs affect census under-coverage only via strata, so  after conditioning  on strata, conditioning on TAs becomes unnecessary. Similarly, we write $\boldXind = \boldx^{\mathrm{ind}}$ instead of $\boldX = \boldx$ in the first element in the integral in (\ref{eq:prediction2}) because conditioning on $\alpha^{\mathrm{hh}}$ means we do not need to condition on $\boldXhh$ and $\boldXpsu$.  The normal density for the household effects in the integrand in (\ref{eq:19}) follows from the model equations (\ref{eq:hh}) and (\ref{eq:psu}) since, by the mixture property of the normal distribution \citep[p577]{BDA3}, we have
\begin{align*}
p(\alpha^{\mathrm{hh}} & \vert \boldX^{\mathrm{hh}}=\boldx^{\mathrm{hh}},\boldX^{\mathrm{psu}}=\boldx^{\mathrm{psu}}, \mathrm{strat}=s, \boldxi) \nonumber \\
& = \int p(\alpha^{\mathrm{hh}} \vert \alpha^{\mathrm{psu}},\boldX^{\mathrm{hh}}=\boldx^{\mathrm{hh}},\boldX^{\mathrm{psu}}=\boldx^{\mathrm{psu}}, \mathrm{strat}=s, \boldxi)p(\alpha^{\mathrm{psu}} | \boldX^{\mathrm{psu}}=\boldx^{\mathrm{psu}},\alpha_{s}^{\mathrm{strat}}) \, \mathrm{d}\alpha^{\mathrm{psu}}  \nonumber \\
& = \int \mathcal{N} \left (\alpha^{\mathrm{hh}} | \mu + \boldx^{\mathrm{hh}'}\boldbeta^{\mathrm{hh}} + \alpha^{\mathrm{psu}}, \sigma_{\mathrm{hh}}^{2} \right ) %
\mathcal{N} \left (\alpha^{\mathrm{psu}} | \boldx^{\mathrm{psu}'}\boldbeta^{\mathrm{psu}} + \alpha_{s}^{\mathrm{strat}},\sigma_{\mathrm{psu}}^{2} \right ) \, \mathrm{d}\alpha^{\mathrm{psu}} \nonumber \\
& = \mathcal{N} \left (\alpha^{\mathrm{hh}} \vert \mu + \boldx^{\mathrm{hh}}\boldbeta^{\mathrm{hh}} +
\boldx^{\mathrm{psu}}\beta^{\mathrm{psu}} + \alpha_s^{\mathrm{strat}}, \sigma_{\mathrm{hh}}^2 + \sigma_{\mathrm{psu}}^2 \right ). 
\end{align*}
$\Pr(\mathrm{strat}=s \vert \mathrm{TA}=t)$ is estimated using an occurrence matrix constructed using the same data as $\boldW$, that is the official census file, which augments the census respondent file with administrative records.

The integral in (\ref{eq:19}) produces predicted coverage probabilities that are marginalised with respect to household and PSU effects. That is, they are not predictions that are relevant to particular households, but are expectations over the distribution of household effects among households with covariate values $\boldx^{\mathrm{hh}}$ in PSUs with covariates $\boldx^{\mathrm{psu}}.$ An alternative, conditional prediction, could be obtained by setting the household and PSU effects to zero (or some other value) but such predictions are tied to households and PSUs with the specified effect and would not be appropriate for application to the census file for which the desired notion is that of an unknown household with particular household covariate values in an unspecified PSU with particular PSU covariate values. Further discussion on the marginal and conditional predictions can be found in \citet{skrondal2009prediction} and \citet{pavlou2015prediction} and some more details on the derivation of (\ref{eq:19}) are given in section B of the Supplementary Material. In our application we use Monte Carlo integration to approximate the integral in (\ref{eq:19}).

\section{Results
\label{sec:results}}

\subsection{Using mixed predictive checks to assist with model assessment \label{res:ppc}}

Models were run with three HMC chains of sufficient length (11 000-12 000 iterations) to ensure convergence. Stan run times with parallel chains were 8.0 hours for the initial model (model 1), and 17.9 hours for the model with household covariates (model 2, see section \ref{res:hh_covar}). After discarding the first half of each chain as warm-up period, the $\hat{\mathrm{R}}$ convergence diagnostic \citep[pp 284-285]{BDA3} was less than 1.01 for all monitored parameters for both models. We assess the quality of the models using posterior predictive checking focusing on marginal predictions for two different groupings: demographic categories formed by all binary demographic covariates, and TAs. Results for all checks performed on two models are presented in Figure \ref{fig:ppc_demog} for predictions on demographic groupings, and in Figure \ref{fig:ppc_ta} for TA-level predictions. Note that only the first 42 TAs of the North Island are shown in Figure \ref{fig:ppc_ta} (see Supplementary Material Figure S2 for all other TAs). For the first predictive check, we use each sample from the joint posterior distribution of parameters to replicate the PES data under the logistic model described in equations (\ref{eq:bern})-(\ref{eq:logit}). We compare the 1000 simulated datasets with the observed data. We summarise the aggregated undercount distributions from simulated datasets using 90\% posterior predictive intervals and assess whether observed undercounts fall within these intervals (Figures \ref{fig:ppc_demog}a and \ref{fig:ppc_ta}a, top intervals). This self-consistency check allows us to confirm that the model fits the data: all observed aggregated undercounts fall within the 90\% posterior predictive intervals from simulated data, for both demographic and geographical groupings.

The PES model is designed to predict under-coverage for census records. Census records can be considered as ``new observations" that we need the model to output predictions for. We therefore need to assess not only the fit but also the predictive ability of the model when confronted with new observations. This is especially important as these observations do not fit into the hierarchy of households, PSUs, and strata that was solely defined to account for the PES sampling design. Therefore, we need to determine how good the model is at estimating under-coverage probabilities for census records, which are characterised by the same TA and demographic information as PES records but are for the most part not included in the households and PSUs selected for PES. This can be done using mixed predictive checking, whereby predictions are performed for new individuals (outside of the PES sampling frame) with exactly the same demographic and group-level predictors as PES individuals \citep{gelman1996posterior}. In our case, it amounts to applying equation (\ref{eq:19}) to all PES records, drawing from the Bernoulli process to simulate the under-coverage indicator, and aggregating the results to the same groupings as previous posterior predictive checks (TA and demographic categories). The results are displayed as light grey 90\% credible intervals (bottom interval) on Figures \ref{fig:ppc_demog}a and \ref{fig:ppc_ta}a. With results aggregated to demographic groups, the model shows a substantial misfit for five out of the 14 most common demographic groups, which is a higher proportion than the 10\% roughly expected under the assumption that the model is adequate. The TA grouping also shows widespread misfit, with almost a third of all TA under-coverage counts lying outside of the predicted 90\% credible intervals.

When a misfit is observed after integration of the sampling design parameters, it is useful to investigate what level of the hierarchy causes the problem, especially in models with more than two levels. In our case, unaccounted for variation could be present at the individual, household, PSU, or stratum level. To assess the problematic level, we compute mixed predictive checks where some but not all of the grouping levels in the hierarchy are integrated. We first perform predictions from the PES data considering that the stratum and PSU of each individual are known but the household is new, therefore sampling the varying intercept from the population distribution for households with their given covariates. For an individual in PSU $p$, in a household with household covariate values $\boldx^{\mathrm{hh}}$ and with individual covariate value $\boldxind$,  this means calculating the following under-coverage probability:

\begin{align*} 
  q^{\mathrm{psu}}(\boldxind,\boldx^{\mathrm{hh}},\boldxi,\alpha_{p}^{\mathrm{psu}}) &= \int \mathrm{expit}( \alpha^{\mathrm{hh}} 
    + \boldx^{\mathrm{ind}\prime}\boldbeta) \mathcal{N}(\alpha^{\mathrm{hh}}
    \vert \mu + \boldx^{\mathrm{hh}'}\boldbeta^{\mathrm{hh}} +
    \alpha_{p}^{\mathrm{psu}},\sigma_{\mathrm{hh}}^2) \, \mathrm{d}\alpha^{\mathrm{hh} }.
\end{align*}

The results are displayed as dark grey 90\% credible intervals (second from the top) on Figures \ref{fig:ppc_demog}a and \ref{fig:ppc_ta}a.
We repeat the procedure where both households and PSUs are new, which in practice consists in sampling the intercept from the population distribution for households, integrated over PSUs. In this case, the under-coverage probability in stratum $s$ is calculated as follows:
\begin{align*} 
  q^{\mathrm{strat}}(\boldxind,\boldx^{\mathrm{hh}}, & \boldx^{\mathrm{psu}},\boldxi)  = \nonumber \\ 
   & \int \mathrm{expit}(\alpha^{\mathrm{hh}} + \boldx^{\mathrm{ind}\prime}\beta) \mathcal{N}(\alpha^{\mathrm{hh}} \vert \mu + \boldx^{\mathrm{hh}'}\boldbeta^{\mathrm{hh}} +
 \boldx^{\mathrm{psu}'}\boldbeta^{\mathrm{psu}} +\alpha_{s}^{\mathrm{strat}},\sigma_{\mathrm{hh}}^2 + \sigma_{\mathrm{psu}}^2) \, \mathrm{d}\alpha^{\mathrm{hh}}.
\end{align*}

The results of these predictions are displayed as grey 90\% credible intervals (third from the top) on Figures \ref{fig:ppc_demog}a and \ref{fig:ppc_ta}a.

\subsection{Adding household covariates improves the model for demographic groups \label{res:hh_covar}}

\begin{figure}[h]
\centering
\includegraphics[width=\textwidth]{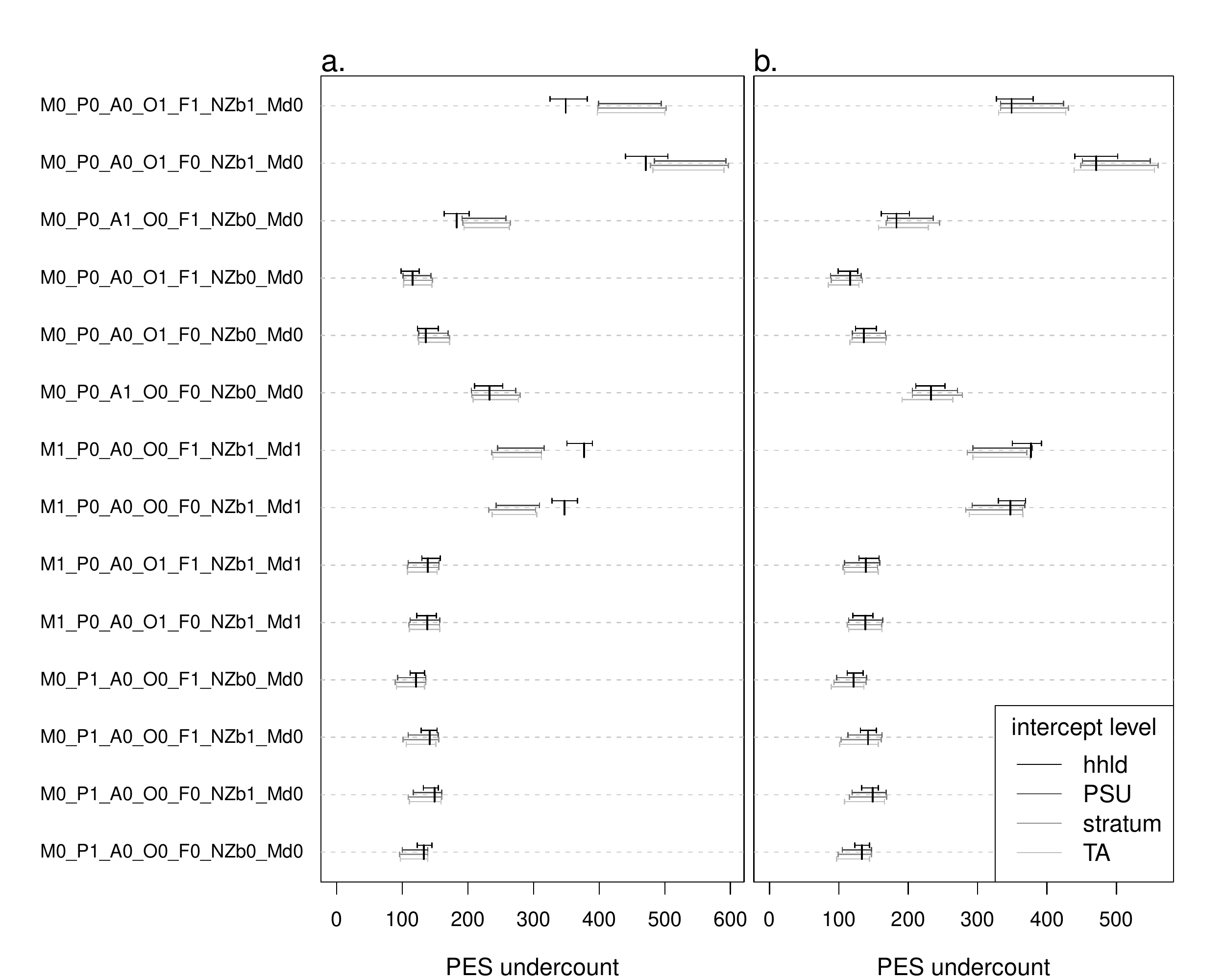}
\caption{Posterior predictive checks (PPCs) for model 1 (without household covariates, left) and model 2 (with household covariates and their interactions, right). Only the 14 most common demographic categories are represented, out of 81 demographic categories, all geographic areas pooled. The vertical black bars represent the PES observed counts for the categories displayed on the left axis. PPC results for different levels of integration are represented by horizontal 90\% credible intervals. For each category, the top interval corresponds to PPCs performed on the raw output from the logistic regression, and lower intervals (different shades of grey) correspond to PPC results after Monte Carlo integration of household parameters, household and PSU parameters, and household, PSU, stratum parameters, respectively. The labels on the left are to be interpreted as a combination of demographic variables (letters) with whether or not the category comprises individuals corresponding to the demographic variable (0=no, 1=yes). M=Māori, P=Pacific, A=Asian, O=Other, F=Female, NZb = New-Zealand born, Md = Māori descent. For instance the top category corresponds to females of Other ethnicity only, who were born in New Zealand and are not of Māori descent.}
\label{fig:ppc_demog}
\end{figure}

\begin{figure}[!htb]
\centering
\includegraphics[height=.9\textheight]{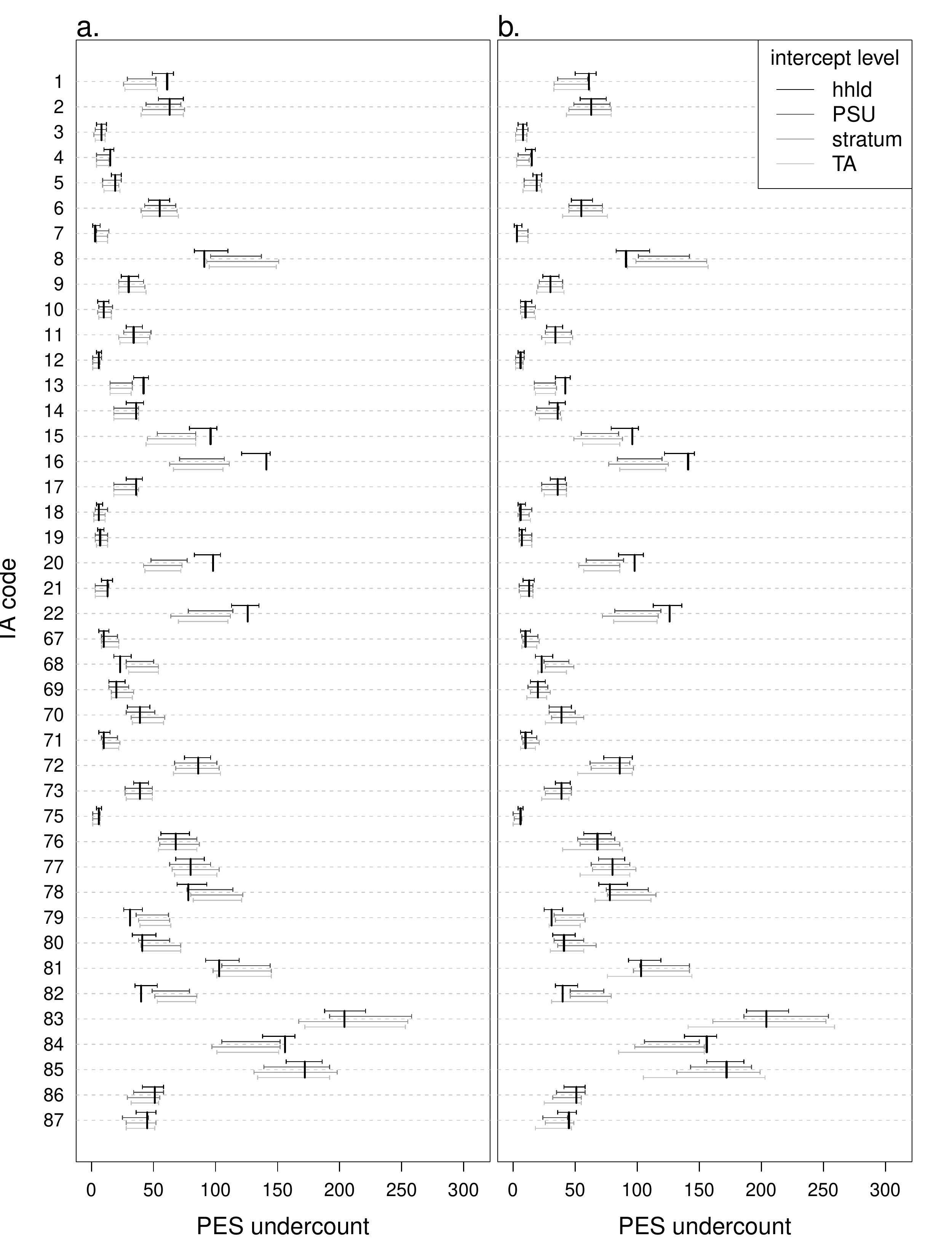}
\caption{Posterior predictive checks (PPCs) for model 1 (without household covariates, left) and model 2 (with household covariates and their interactions, right). Undercounts by TAs are represented for the first 42 TAs of the North Island, all demographic categories pooled. See Figure \ref{fig:ppc_demog} caption for further details.}
\label{fig:ppc_ta}
\end{figure}

The four performed predictive checks, with integration occuring at different grouping levels, clearly show that a major misfit arises when predicting under-coverage of individuals in new households. 
This misfit is visible for both demographic groupings (fig.\ref{fig:ppc_demog}a) and TA groupings (fig.\ref{fig:ppc_ta}a), but does not grow larger when predictions are calculated with unknown PSUs and/or strata. This suggests that the model did not estimate an adequate population distribution for households.
Further graphical investigations (not shown) suggest some dependence between the varying intercept and individual covariates. As noted in section \ref{subsec:modelspec}, the logistic model described in (\ref{eq:bern})-(\ref{eq:ta}) assumes independence between intercept and individual-level predictors. Following \citet[p. 506]{gelman2006data}, we address this inconsistency by aggregating the individual covariates suspected to cause the dependency to the hierarchical level in question, and introduce the new variables as group-level covariates. We first test different ways of aggregating demographic covariates (ethnicity, age, sex, and New-Zealand born) at the household level. For continuous variables, it makes sense to average values across individuals in a group. However, several ways to aggregate categorical variables at the group level can be considered. 
For each individual categorical covariate, we test the following aggregation methods: (i) binary variable indicating presence/absence of household occupants with the demographic characteristic, (ii) continuous variable of proportion of household occupants with the demographic characteristic, and (iii) binary variable indicating a majority of household occupants with the demographic characteristic. 
We find that overall model performance is best when using (i) for all group-level covariates. We therefore only present results with these covariates (see Table \ref{table:covariates} for the final list of household covariates and their interactions). 
We apply the same four predictive checks as we applied to the original model, and present the results in Figures \ref{fig:ppc_demog}b and \ref{fig:ppc_ta}b.
 
Introducing household covariates and their interactions considerably improves the fit of predicted undercounts of demographic groups to the observed data (fig. \ref{fig:ppc_demog}), with only one observed value sitting just outside the 90\% credible intervals from the TA-level prediction. However, the modification only partially improves the fit to TA counts (fig. \ref{fig:ppc_ta}). While the model including household covariates fits most TAs well, 7 TAs on Figure \ref{fig:ppc_ta} still show predictive credible intervals that contain the true value when simple posterior predictive checks are performed but do not encompass it when performing any of the mixed predictive checks. For these TAs, the estimated household population distribution seems wrong, and we hypothesise that some unknown household characteristics cause unaccounted-for heterogeneity. Consistent  with this hypothesis, between household variation is the largest  component of unexplained variation and did not shrink after the inclusion of household level covariates (Table \ref{table:variances}). 

\begin{table}[ht]
\centering
\begin{tabular}{lllllll}
\cline{2-7}
              & \multicolumn{3}{l}{model 1} & \multicolumn{3}{l}{model 2} \\ \cline{2-7} 
              & 2.5\%  & 50\%  & 97.5\%  & 2.5\%  & 50\%  & 97.5\%  \\ \hline
$\sigma^{\mathrm{hh}}$   & 4.10  & 4.35  & 4.62   & 4.18  & 4.45  & 4.74   \\
$\sigma^{\mathrm{psu}}$  & 0.29  & 0.76  & 1.05   & 0.08  & 0.63  & 0.98   \\
$\sigma^{\mathrm{strata}}$ & 0.01  & 0.18  & 0.45   & 0.01  & 0.15  & 0.42   \\
$\sigma^{\mathrm{ta}}$   & 0.01  & 0.13  & 0.43   & 0.01  & 0.12  & 0.43   \\ \hline
\end{tabular}
\caption{Marginal posterior quantiles of variance parameters for model 1 (without household-level deomgraphic covariates) and model 2 (with household-level demographic covariates)}
\label{table:variances}
\end{table}

\subsection{Complementary cross-validation tests}

Posterior predictive checks give insight into the fit of the model to the data. They are a first ``sense-check" of an analysis. The three additional mixed predictive checks, simulating data with the same covariates as the data but different hierarchical groups, constitute a step further in assessing not only the fitness but also the predictive power of the model, and where its limitations lie. However, these checks still use the exact same covariate values in the simulated datasets as in the observed datasets. One way to further determine the predictive power of the models is cross-validation. For all tested models, and to assist with model selection, we calculated approximate leave-one-out-cross-validation scores using Pareto-smoothed importance sampling (PSIS). We show the results for the two main models presented in section \ref{res:ppc} and \ref{res:hh_covar} in Table \ref{table:looic}. Lower leave-one-out cross-validation scores (or their importance sampling approximation, LOOIC) indicate a lower out-of-sample deviance, and therefore more accurate predictions to new data. Model 2, with additional household covariates, has a lower LOOIC value than model 1, although the difference is of the order of one standard error. While LOOIC is the most appropriate measure of predictive accuracy for complex hierarchical models, the Pareto-k diagnostic values for both models but especially model 2 suggest the error in the LOO approximations might be high and the LOOIC values might understate predictive accuracy \citep{vehtari2017practical}. This is typical of flexible hierarchical models where some groups have very few observations. Both model 1 and 2 are like this: the lowest level of the hierarchy, households, often contains only one or two observations (individuals).

\begin{table}[h!]
\begin{adjustbox}{max width=\textwidth}
\begin{tabular}{cccccccc}
\hline
\multirow{2}{*}{\textbf{model}} & \multirow{2}{*}{\textbf{$elpd_{loo}$}} & \multirow{2}{*}{\textbf{$p_{loo}$}} & \multirow{2}{*}{\textbf{LOOIC}} & \multicolumn{4}{c}{\textbf{Pareto-k distribution}} \\ \cline{5-8}     \\
                &                   &                 &                 & (-$\infty$,0.5{]} & (0.5,0.7{]} & (0.7,1{]}  & (1,$\infty$)  \\ 
w/o hh covar.     & -5747.1 (100.6)           & 2858.6 (60.3)          & 11494.3 (201.2)         & 65.1\% & 24.0\% & 9.9\% & 0.9\% \\
w/ hh covar.    & -5639.8 (99.4)           & 2814.0 (59.7)          & 11279.7 (198.8)         & 62.1\% & 25.5\% & 11.3\% & 1.1\% \\ \hline
\end{tabular}
\end{adjustbox}
\caption{Out-of-sample deviance diagnostics for the model without household covariates (w/o hh covar.) and with household covariates (w/ hh covar.). $elpd_{loo}$: expected log pointwise predictive density. $p_{loo}$: effective number of parameters. LOOIC: Pareto-smoothed importance sampling leave-one-out cross-validation approximation. Values in brackets correspond to standard error estimates. The Pareto-k distribution section bins estimates of importance of all data records into categories ordered by decreasing quality.}
\label{table:looic}
\end{table}

\subsection{Standardised estimates \label{res:stand}}

The PES model output gives estimates of $p_{under}(\boldv, \boldxi)$ and gives us insight into how different geo-demographic groups respond to census. From a demographer's point of view and for the sake of planning future censuses, it is also valuable to know what factors are actually driving non-response patterns. For instance, in a TA with a high estimated census undercount, it can be of interest to know if non-response is due to the demographic composition of the TA, or if there are there intrinsic difficulties associated with operating a large-scale survey in this area. If demographic effects are predominant, then we can assume non-response is driven by behavioural patterns in the respondents, whereas area effects would suggest potential issues with incomplete address registers or other operational pitfalls.
Insight into the relative impact of the  different covariates on census  coverage can be gained  by calculating under-coverage probabilities across the categories of the variable of  interest for a standardised distribution of all other covariates. For instance, one can obtain area-level estimates where differences due to their demographic composition are statistically removed, leaving only differences pertaining to intrinsic area characteristics. The same standardisation logic can be applied to other variables, for instance one can obtain estimates by ethnicity, standardising areas and all other demographic covariates.
Following the example of TA-level standardised estimates, we can define, for a given TA $t$:
\begin{equation}
p_{under}^{\mathrm{std}}(t,\boldxi) = \sum\limits_{\boldx} p_{under}(\boldx,t,\boldxi)\Pr\textsuperscript{std}(\boldX=\boldx \vert \boldxi), \label{eq:punderstd}
\end{equation}
where $\Pr\textsuperscript{std}()$ refers to the covariate probabilities from some standard distribution. Note that the standard distribution is allowed to depend on the model parameters. This is not usual but suits our situation because a natural choice of standard population is the corrected version of the census file, based on the under-coverage probabilities, estimated from the model. Thus, if the inverse under-coverage probability for the $i^{th}$ census record corresponding to a particular setting of parameter values $\boldxi$ is $w_i(\boldxi)$ we can define $\Pr\textsuperscript{std}(\boldX=\boldx \vert \boldxi)$ as
\begin{align}
\Pr\textsuperscript{std}(\boldX=\boldx \vert \boldxi) = \frac{\sum\limits_{i:\boldX_i=\boldx} w_i(\boldxi)}{\sum\limits_i w_i(\boldxi)}, \label{eq:Prstd}
\end{align}
where the summations are over records in the census file. With the standard probabilities defined as in (\ref{eq:Prstd}), standardised under-coverage probabilities can be obtained for each TA and repeating this for each draw from the posterior for $\boldxi$ will produce a sample from the joint posterior for the standardised under-coverage probabilities by TA. Credible intervals and other summaries, including for contrasts between TAs, can be computed from the posterior sample. The standardised coverage probability, given by (\ref{eq:punderstd}) can be contrasted with the marginal TA under-coverage probability which is
\begin{align}
\Pr(Y=1 \vert TA=t, \boldxi) = \sum\limits_{\boldx} p_{under}(\boldx,t,\boldxi)\Pr(\boldX=\boldx \vert TA=t, \boldxi). 
\label{eq:margucov}
\end{align}

Comparing (\ref{eq:margucov}) and (\ref{eq:punderstd}) it can be seen that they differ only in the covariate distribution, with the standardised probabilities using the covariate distribution of the chosen standard population in place of the TA-specific covariate distribution used to obtain the marginal coverage probability. By definition, the standardised probabilities are all based on the same covariate distribution, so differences in standardised TA under-coverage probabilities reflect genuine geographic differences in census under-coverage.

\begin{figure}[h]
\centering
\begin{adjustbox}{max height={0.9\textheight}}
\includegraphics[width=\textwidth]{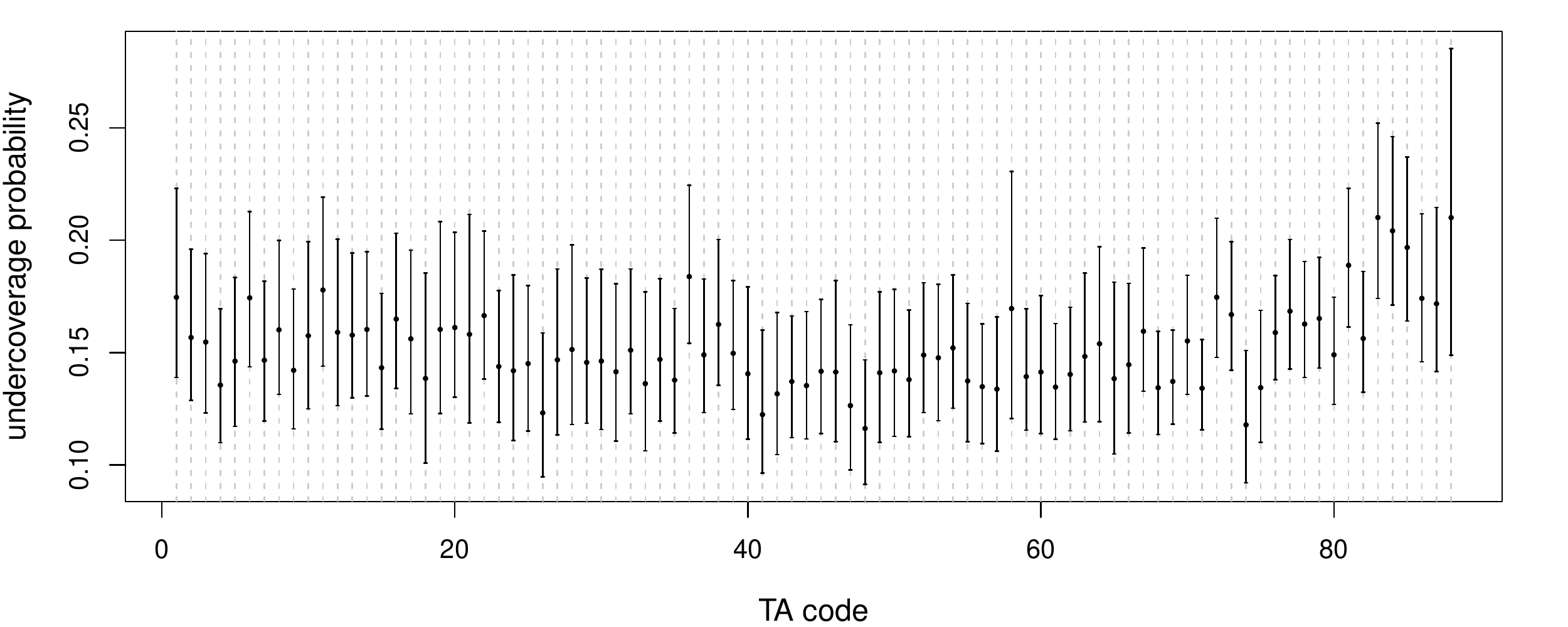}
\end{adjustbox}
\caption{Standardised estimates of under-coverage probabilities for each TA. Points correspond to posterior medians and error bars correspond to 90\% credible intervals.}
\label{fig:stand_ta}
\end{figure}

\begin{figure}[h]
\centering
\begin{adjustbox}{max width={\textwidth}}
\includegraphics[width=\textwidth]{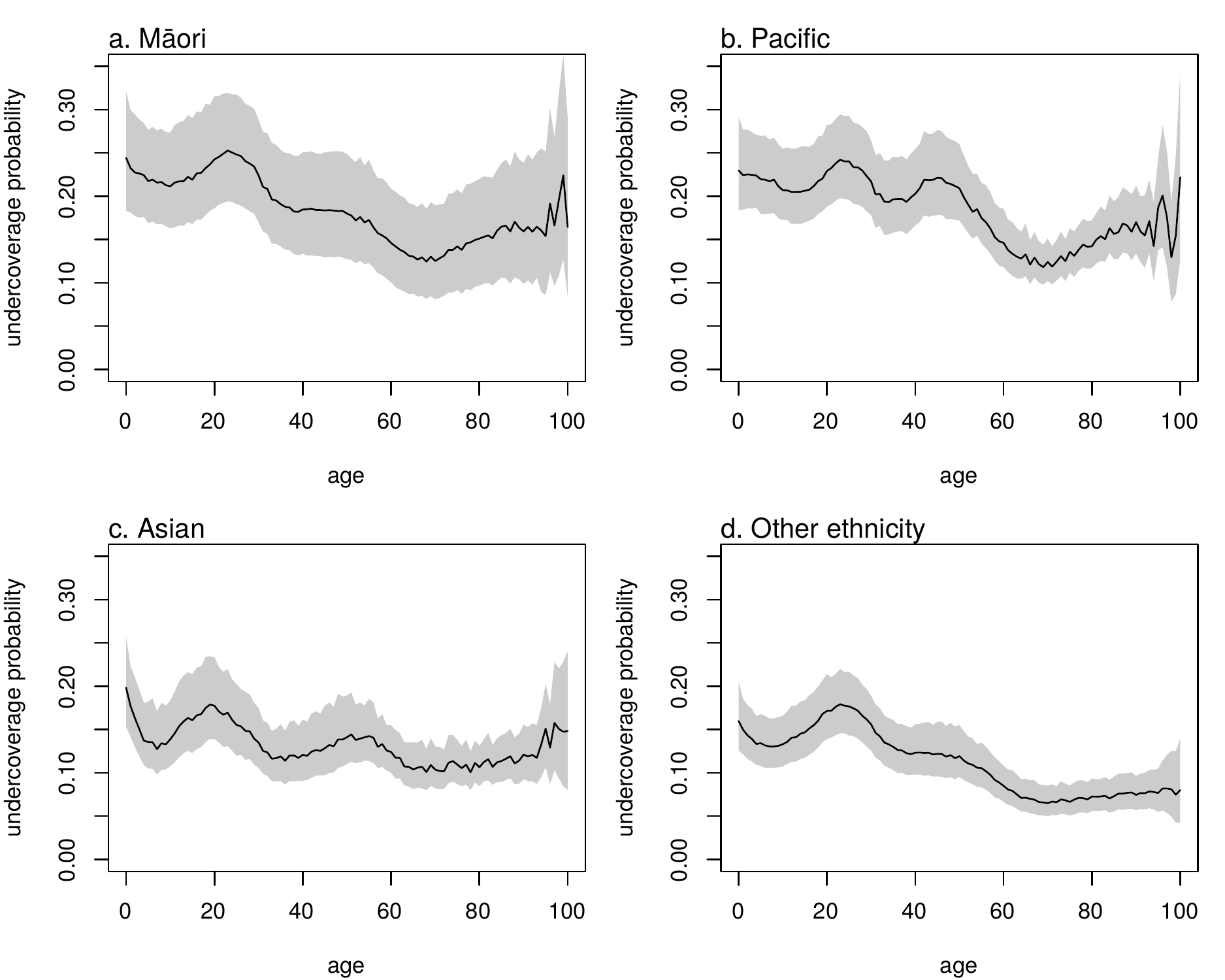}
\end{adjustbox}
\caption{Standardised estimates of under-coverage probabilities by age, for each ethnicity. Black lines correspond to posterior medians and grey shading corresponds to 90\% credible intervals.}
\label{fig:stand_age_eth}
\end{figure}

We focus on standardised under-coverage probability estimates for TAs and for age-sex profiles, using the output from the model with household covariates (model 2). Figure \ref{fig:stand_ta} shows results across TAs. Although uncertainty bounds are wide and overlap among most TAs, some TAs seem to have higher non-response probabilities than the majority, all else being equal. As our mixed predictive checks have identified some inaccuracies in the under-coverage predictions for some TAs (section \ref{res:hh_covar}), we are cautious about drawing conclusions on our TA-level standardised estimates. Figure \ref{fig:stand_age_eth} shows that \Maori and Pacific people generally have higher non-response levels than other ethnic groups. People in their twenties have the highest non-response levels across all ethnic groups. A secondary under-coverage peak around age 50 is also visible, although it is more pronounced in Pacific people than in other ethnic groups.

\subsection{Effects of coverage adjustment on census counts \label{res:car}}

As stated in section \ref{subsec:modelspec}, census counts by demographic and geographic categories are corrected using the posterior values for $p_{under}(\boldv_{i},\boldxi)$, following equation (\ref{eq:car}). Although producing adjusted counts is out of scope for this paper, it is useful to provide a sense of the scale of the correction to the census counts implied by the estimated under-coverage probabilities. Assuming model 2 is chosen, we find that overall average undercoverage of census responses is 10.9\%. For a hypothetical census dataset comprising four million responses this would mean adding about 490,000 individuals to the census respondent population. However, some demographic groups are better represented in census than others. We find that census under-coverage in young male adults of Pacific and \Maori populations can reach 35\%. This means that the population for such categories is about 1.5 times larger than in the census responses, though in the official census file this effect is offset by the inclusion of adminstrative records.

\section{Discussion
\label{sec:discuss} }

\subsection{Posterior predictive checks help understand and select the multilevel model
\label{subsec:disc_ppc} }

Through the present analysis, we have illustrated how census coverage can be quantified using a modelling approach to the analysis of complex survey data, thereby allowing insights at high levels of granularity in demographic and geographic attributes. To estimate under-coverage of the New Zealand 2018 census, we have fitted a binomial model with four nested geographic levels reflecting the complex sampling design of the post-censal survey. Especially, we have shown that multilevel models are a promising approach to analyse survey data with complex sampling designs.
We reiterate that our results are experimental results that reflect a different analysis to the one used to output official 2018 census coverage estimates \citep{PES2018}.

We have also illustrated how performing extensive posterior predictive checking can assist in model selection in the case of a complex multilevel model structure. Combining posterior predictive checking, mixed predictive checking and cross-validation allowed us to assess the fit and predictive limitations of competing models as well as identifying aspects of the models that require modifications. Especially, performing mixed predictive checking at all levels of the hierarchy allowed us to identify the lack of fit at a specific level (here, the household level). We could assess the improvement associated with the subsequent addition of household-level covariates using additional mixed predictive checks and comparing cross-validation results across models. 

Even after the addition of household-level covariates, posterior predictions did not always fit under-coverage data at the TA level. Some TA under-coverage counts were over-estimated, with 10 (of 88) TAs having PES-observed census under-coverage counts below the predicted 90\% credible intervals. Other TA counts were under-estimated, with 12 TAs having observed under-coverage counts over the predicted 90\% credible intervals (fig. \ref{fig:ppc_ta} and S2). Mixed posterior predictive checks at each level of geographic parameter integration allowed us to attribute most of this misfit to the household level. The lack of fit is unlikely to be related to the demographic attributes of household occupants, as most of these attributes have been accounted for as household-aggregated level variables, and posterior predictive checks for demographic groups show no apparent bias. Subsequent investigations have failed to identify commonalities between TAs with similar misfit patterns. 

The only noticeable pattern in this result is the relationship between observed under-coverage proportion of a TA and the direction of the estimation error: TAs which tend to be under-estimated are the ones with high observed under-coverage, whereas over-estimated TAs tend to have a low observed under-coverage proportion. Taken at face value this result suggests the model may be over-regularising more extreme estimates. In multilevel modelling, one expects small hierarchical groups with extreme observations to have predictions shrunk towards their expected value under the model. As the number of levels increases, we can expect shrinkage to increase too for predictions made at higher levels. However, if over-shrinkage, \textit{per se}, was the main reason for the TA-level misfit, we would expect the issue to primarily affect smaller areas, whereas several of the TAs that the model fits poorly are large urban areas with a relatively large PES sample size. Further, if over-shrinkage was the primary reason for the lack of model fit in some areas, we would expect predictions to gradually show more shrinkage as we move from household-level predictions towards TA-level predictions, instead of a single jump from adequate fit at the household level to misfit for prediction at all other levels. We experimented with replacing the normal distributions for household, PSU and stratum effects by the heavier tailed t distribution with three degrees of freedom, but this had no impact on results. If the lack of fit in some TAs was due to the normal models tending to over-shrink extreme values we would have expected to see some improvement in fit when $t_{3}$ priors were adopted. The most plausible explanation for the lack of fit in some TAs is that one or several important factors related to geography were not included in the model. If this explanation holds, it follows that estimates are, in some cases, being shrunk towards expectations that do not exhibit the appropriate amount of geographic variation because a geographically varying covariate has been omitted from the model. In this specific sense, the estimates for some TAs may be exhibiting the effects of over-shrinkage. As noted above, given our posterior predictive checking results, the omitted covariate(s) seems likely to be a household-level variable, such as an aspect of dwelling construction (e.g free-standing versus in an apartment block) that varies by area and is related to census coverage (e.g. census enumeration may be more difficult in apartment blocks). A natural next step for future model improvement would involve attempting to identify the missing covariate(s). If the missing covariates cannot be identified or sourced, a potential alternative is to specify the problematic group distribution as a mixture of several distributions. This may allow the model to recover unobserved categories within groups and improve model fit. However moving to mixture distributions at one or more of the model levels introduces additional computational complexity.

\subsection{Individual demographic characteristics drive census coverage patterns
\label{subsec:disc_stand} }

Standardised estimates give insight into the role played by different demographic and geographic attributes in driving coverage differences between groups. Though common practice in epidemiology and demography, the use of standardisation to adjust for differences in covariate distributions has been less common in official statistics. In our case it provided a simple way to present comparative results from a complex model. Comparing Figures \ref{fig:stand_ta} and \ref{fig:stand_age_eth} suggests high census under-coverage patterns are in general driven more by individual demographic attributes than by geographic ones. For instance, \Maori and Pacific people as well as people in their early twenties tend to respond to census less than other demographic groups, independently of where they live. Although most TAs do not seem to intrinsically drive census under-coverage, clusters of TAs with higher under-coverage propensity can be identified. In this case standardised estimates can be used in the planning of future census operations. For instance, incentivisation and follow-up efforts could be allocated more heavily in TAs where under-coverage propensity has historically been high. 

\subsection{Design-based vs. model-based approach
\label{subsec:disc_desvsmod} }

The most common approach to analysing complex surveys has traditionally been through a design-based method, where individual sampling weights are calculated from the sampling frame and subsequently adjusted for non-response. This approach has limitations when survey non-response is difficult to track. For instance, we do not know the number of occupants in non-responding households nor the number of non-respondents in a responding household. Sampling weights are often adjusted for non-response using benchmark population data to ensure that weighted sample distributions are close to known population distributions. However, for PES, such benchmark population data is not available, because PES is used in conjunction with census to estimate a new benchmark population. A further challenge to the application of design-based methods in PES is the absence of an accurate count of dwelling numbers by PSU at the time of the PES fieldwork, which complicates the computation of selection probabilities and hence sampling weights. Moreover, a sample size of about 15,000 households does not allow precise design-based estimates at the required level of geographic and demographic disaggregation. In this regard, the modelling approach seems natural, especially when geographical attributes are treated in a hierarchical fashion. Multilevel modelling facilitates pooling of information across areas and is desirable for small area and small domain estimation problems. 

Modelling of survey data is, of course, possible from a design-based perspective, though there appear to be efficiency gains through explicitly modelling the survey design structure rather than dealing with the impact of survey design through sampling weights \citep{lumley2017fitting}. Design-based multilevel modelling is challenging, because the pseudo-likelihood methods commonly used for design-based fitting of single-level models are more difficult  to apply in the case of multilevel models. Pseudo-likelihood estimates of multilevel models are potentially sensitive to the scaling of survey weights, even when design and analysis clusters are identical \citep{rabe2006multilevel}. Methods based on pairwise composite likelihood are a promising alternative to pseudo-likelihood methods for fitting design-based multilevel models but require knowledge of joint selection probabilities \citep{rao2013weighted,yi2016weighted}. In the PES analysis the geographic clusters of analytical interest are the TAs, which were not part of the sample design and this further complicates the application of design-based methods to multilevel modelling \citep{lumley2017fitting}.

\subsection{Mitigating the ignorable inclusion  assumption
\label{subsec:disc_indep}}

One of the fundamental assumptions of the PES model is the independence between inclusion in PES and inclusion in census, conditional on design features and covariates included in the model. This means that the list of dwelling addresses used for census and the PES sampling frame need to be built independently, a requirement sometimes difficult to satisfy. Another challenge to the conditional independence assumption is respondent behaviour. For instance, a respondent's negative experience with census might influence whether they open the door to PES interviewers. Such behaviour would lead to non-ignorable non-response and complicate the analysis by requiring that the model for inclusion in PES be explicitly formulated and included in the model fitting. \citet{pfeffermann2006multi} develops a conditional likelihood approach to incorporating non-ignorable non-response  in multilevel modelling of survey data. Extending the PES model to deal with non-ignorable non-response may be a worthwhile direction for future development of the model. In Figure \ref{fig:dag}, this would result in additional edges between one or more of the inclusion indicators and the coverage indicator, illustrating the need to  explicitly specify the model for inclusion and to  estimate the inclusion model jointly with  the coverage model. Alternatively, it may be possible to incorporate external information that allows the assumption of conditional independence between PES and census inclusion to be weakened \citep{elliott2000bayesian,Brownonscoverage}.

\section*{Disclaimer}
The views expressed in this paper are those of the authors and should not be taken to represent an official view of their affiliated organisation.

\section*{Acknowldgements}
We thank three annonymous referees and an Associate Editor for thoughtful comments.

\bibliography{bib18_resubmit2}

\end{document}


\maketitle

\newpage

\appendix

\section{Supplementary figures}

\begin{figure}[h!]
\centering
\includegraphics[scale=1]{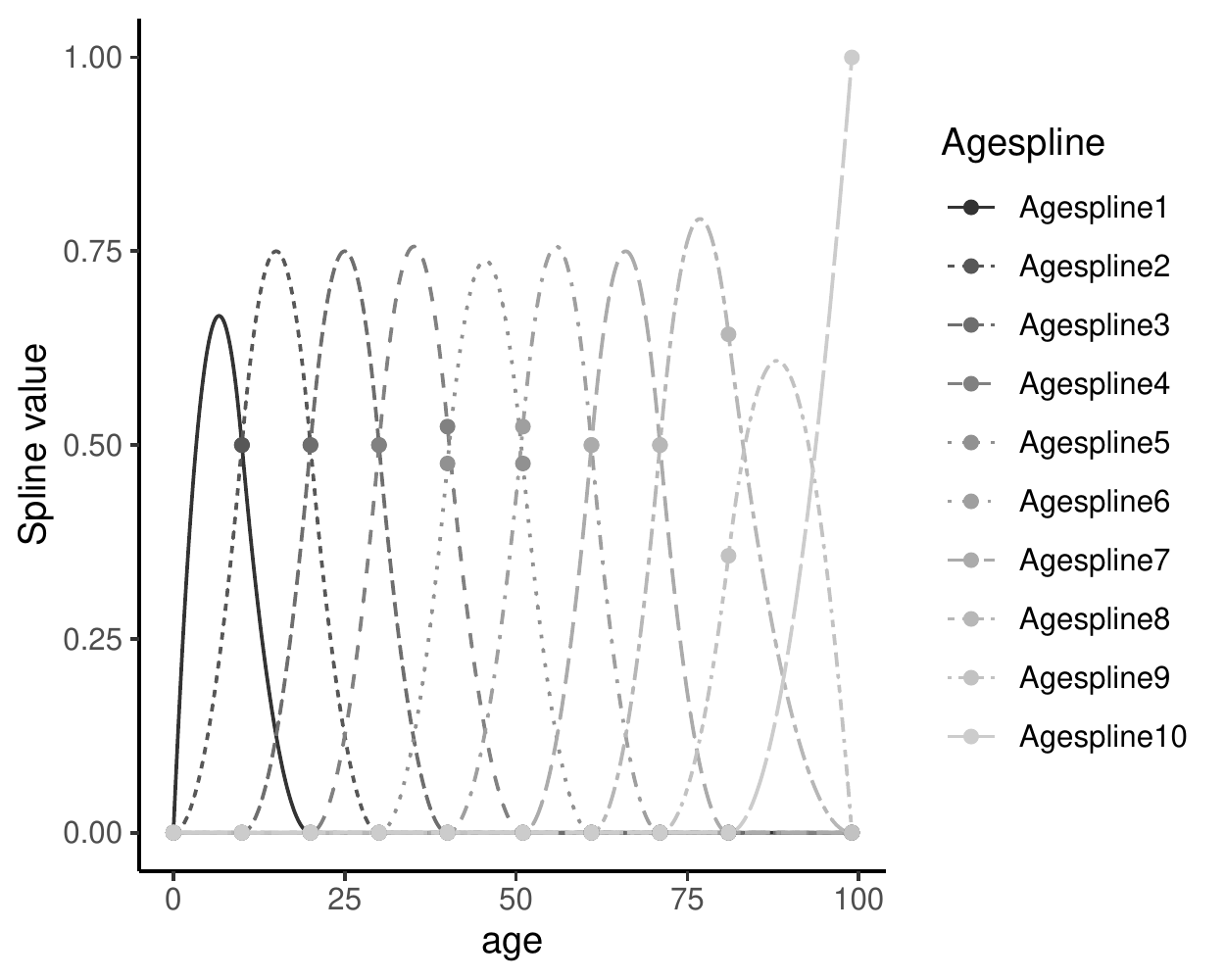}
\caption{Spline basis fitted to PES age variable. Dots show specified knots.}
\label{fig:agesplines}
\end{figure}

\begin{figure}[h!]
\centering
\includegraphics[width=\textwidth]{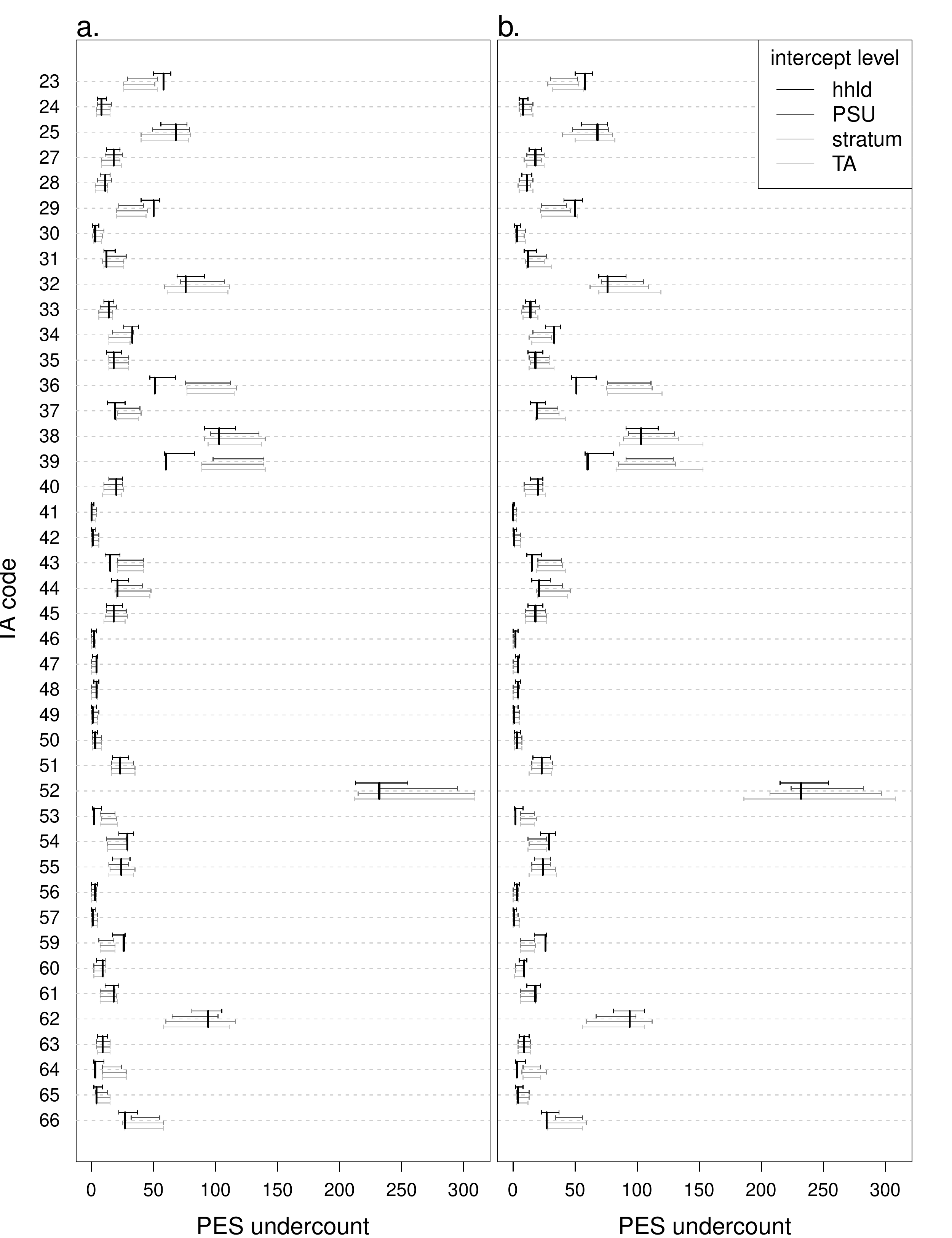}
\caption{Posterior predictive checks (PPCs) for model 1 (without household covariates, left) and model 2 (with household covariates and their interactions, right). Undercounts by TAs are represented for all TAs not represented in figure 3 in the main text, all demographic categories pooled.}
\label{fig:ppc_ta_SI}
\end{figure}

\newpage

\section{Equivalence of predictions obtained from models with household and PSU covariates included either at their natural level or at the individual level}\label{proof_indiv_grp_covar}

In this appendix we verify that the model specification in which household and PSU level covariates are included in the model as covariates at the individual level yields the same predicted probabilities as the model given by (2-7) in which covariates are included at their natural level.
	
	Firstly, we provide some additional background on the equivalence of the model specification given by (2-7) and the specification with household and PSU level covariates included at the individual level of the model. In the latter formulation all individuals in a household or PSU are assigned the covariate values for their household or PSU, in addition to their individual level covariate values. We used this formulation of the model for model fitting.
	
	The household level model (4) implies the household effect for the $h^{th}$ household can be written
	\begin{equation*}
	\alpha_{h}^{\mathrm{hh}} = \mu + \boldX_{h}^{\mathrm{hh}'}\boldbeta^{\mathrm{hh}} + \delta_{h}^{\mathrm{hh}}, 
	\label{eq:althh}
	\end{equation*}
	for $\delta_{h}^{\mathrm{hh}} \sim N(\alpha_{\mathrm{psu}[h]}^{\mathrm{psu}},\sigma_{\mathrm{hh}}^{2} ).$
	Similarly, the PSU level model implies the $p^{th}$ PSU effect can be written
	\begin{equation*}
	\alpha_{p}^{\mathrm{psu}} = \boldX_{p}^{\mathrm{psu}'}\boldbeta^{\mathrm{psu}} + \delta_{p}^{\mathrm{psu}},
	\label{eq:altpsu}
	\end{equation*}
	for $\delta_{p}^{\mathrm{psu}} \sim N(\alpha_{strat[p]}^{\mathrm{strat}},\sigma_{\mathrm{psu}}^{2}),$
	so
	\begin{equation}
	\delta_{h}^{\mathrm{hh}}=\boldX_{\mathrm{psu}[h]}^{\mathrm{psu'}}\boldbeta^{\mathrm{psu}} + \lambda_{h}, \nonumber
	\end{equation} %
	where $\lambda_{h} \sim N(\delta_{\mathrm{psu}[h]}^{\mathrm{psu}},\sigma_{\mathrm{hh}}^{2}).$ Consequently,
	\begin{equation}
	\alpha_{h}^{\mathrm{hh}} = \mu + \boldX_{h}^{\mathrm{hh}'}\boldbeta^{\mathrm{hh}} + \boldX_{\mathrm{psu}[h]}^{\mathrm{psu}'} \boldbeta^{\mathrm{psu}} + \lambda_{h} 
	\label{eq:althh2}
	\end{equation}
and substituting (\ref{eq:althh2}) for $\alpha_{h}^{\mathrm{hh}}$ in the individual level model (3)
	 gives the alternative multilevel model specification
	\begin{align}
	\left [ Y_{hj} | p_{under_{hj}} \right ] & \indep \mathrm{Bernoulli}(p_{under_{hj}}), j= 1,\ldots, N_{h}^{\mathrm{ind}}; \, h=1,\ldots, N_{\mathrm{psu}[h]}^{\mathrm{hh}}, \label{eq:Bernoulli} \\
	\mathrm{logit}(p_{under_{hj}} ) & = \mu + \boldX_{hj}^{\mathrm{ind}'}\boldbeta + \boldX_{h}^{\mathrm{hh}'}\boldbeta^{\mathrm{hh}} + \boldX_{\mathrm{psu}[h]}^{\mathrm{psu}'}\boldbeta^{\mathrm{psu}} + \lambda_{h}, \nonumber \\
	& \, \, \,  j= 1,\ldots, N_{h}^{\mathrm{ind}}; \, h=1,\ldots, N_{\mathrm{psu}[h]}^{\mathrm{hh}}, \label{eq:logitmodel}
	\\
	\left [\lambda_{h}|\delta_{\mathrm{psu}[h]},\sigma_{\mathrm{hh}}^{2} \right ] & \indep \mathcal{N}(\delta_{\mathrm{psu}[h]},\sigma_{\mathrm{hh}}^{2}), h = 1,\ldots N_{\mathrm{psu}[h]}^{\mathrm{hh}}, \label{eq:dwellmodel} \\
	\left [ \delta_{p}^{\mathrm{psu}} | \alpha_{strat[p]}^{\mathrm{strat}},\sigma_{\mathrm{psu}}^{2} \right] & \indep \mathcal{N}(\alpha_{strat[p]}^{\mathrm{strat}},\sigma_{\mathrm{psu}}^{2}), \, p = 1, \ldots, N_{strat[p]}^{\mathrm{strat}}, \label{eq:psumodel} \\
	\left [\alpha_{s}^{\mathrm{strat}} | \boldW,\boldalpha^{\mathrm{ta}},\sigma_{\mathrm{strat}}^{2} \right ] & \indep \mathcal{N} \left (\mathbf{W}_{s}\boldalpha^{\mathrm{ta}},\sigma_{\mathrm{strat}}^{2} \right ), s = 1, \ldots, N^{\mathrm{strat}}, \label{eq:stratummodel}\\
	\left [\alpha_{t}^{\mathrm{ta}}|\boldX_{t}^{\mathrm{ta}},\boldbeta^{\mathrm{ta}}, \sigma_{\mathrm{ta}}^{2} \right ] & \indep t_{3}(\boldX_{t}^{\mathrm{ta}'}\boldbeta^{\mathrm{ta}},\sigma_{\mathrm{ta}}^{2}), t=1,\ldots,N^{ta}. \label{eq:tamodel}
	\end{align}
	
	To obtain the predicted under-coverage probability for a hypothetical individual with covariates $\boldX^{\mathrm{ind}}=\boldx^{\mathrm{ind}},$ residing in a household with covariates $\boldX^{\mathrm{hh}}=\boldx^{\mathrm{hh}},$
	located in a PSU with covariates $\boldX^{\mathrm{psu}}=\boldx^{\mathrm{psu}}$ and in TA, $t$, under the model specification (\ref{eq:Bernoulli}-\ref{eq:tamodel}), we first marginalise over strata to obtain
	\begin{align}
	\Pr(Y=1 | & \boldXind=\boldxind,\boldX^{\mathrm{hh}}=\boldx^{\mathrm{hh}}, \boldX^{\mathrm{psu}}=\boldx^{\mathrm{psu}},\mathrm{TA}=t,
	\boldxi) =\nonumber \\
	& \sum_{s}\Pr(Y=1 | \boldXind=\boldxind,\boldX^{\mathrm{hh}}=\boldx^{\mathrm{hh}},\boldX^{\mathrm{psu}}=\boldx^{\mathrm{psu}},\mathrm{TA}=t,\mathrm{strat}=s,\boldxi )\nonumber \\ 
	& \times \Pr(\mathrm{strat}=s |\mathrm{TA}=t,\boldXind=\boldxind,\boldX^{\mathrm{hh}}=\boldx^{\mathrm{hh}},\boldX^{\mathrm{psu}}=\boldx^{\mathrm{psu}},\xi), \nonumber \label{eq:predict} \\
	\end{align}
	where $\boldxi$ is the vector of coverage model parameters excluding the household effects.
	
	To obtain the probability $\Pr(Y=1 |\boldXind=\boldxind,\boldX^{\mathrm{hh}}=\boldx^{\mathrm{hh}},\boldX^{\mathrm{psu}}=\boldx^{\mathrm{psu}},\mathrm{TA}=t,\mathrm{strat}=s,\boldxi)$ we need to marginalise over the distribution of the household effects, conditional on stratum. 
	Within stratum $s,$ it follows from (\ref{eq:dwellmodel}) and (\ref{eq:psumodel}) the household effects distribution, can be obtained by marginalising over PSU effects, to give 
	\begin{align*}
	p(\lambda_{h} | \alpha_{s}^{\mathrm{strat}},\sigma_{\mathrm{hh}}^{2},\sigma_{\mathrm{psu}}^{2}) & =
	\int \mathcal{N}(\lambda_{h}|\delta_{\mathrm{psu}[h]}^{\mathrm{psu}},\sigma_{\mathrm{hh}}^{2})
	\times \mathcal{N}(\delta_{\mathrm{psu}[h]}^{\mathrm{psu}} |\alpha_{s}^{\mathrm{strat}},\sigma_{\mathrm{psu}}^{2} ) \,
	\mathrm{d}\delta_{\mathrm{psu}[h]}^{\mathrm{psu}} \nonumber \\
	& = \mathcal{N}(\lambda_{h} |\alpha_{s}^{\mathrm{strat}}, \sigma_{\mathrm{hh}}^{2} + \sigma_{\mathrm{psu}}^{2} ), 
	\end{align*}
	by the mixture property of the normal  distribution \citep[p577]{BDA3}.
	Consequently, the stratum-specific predicted under-coverage probabilities in the first term of the summand in (\ref{eq:predict}) can be obtained as 
	\begin{align}
	\Pr(Y=1 |\boldXind=\boldxind, & \boldX^{\mathrm{hh}}=\boldx^{\mathrm{hh}},\boldX^{\mathrm{psu}}=\boldx^{\mathrm{psu}}, \mathrm{TA}=t,\mathrm{strat}=s,\boldxi) = \nonumber \\
	& \int \mathrm{expit} (\mu + \boldx'\boldbeta + \boldx^{\mathrm{hh}'}\boldbeta^{\mathrm{hh}} + \boldx^{\mathrm{psu}'}\boldbeta^{\mathrm{psu}} + \lambda_{h}) \mathcal{N}(\lambda_{h}|\alpha_{s}^{\mathrm{strat}},\sigma_{\mathrm{hh}}^{2}+\sigma_{\mathrm{psu}}^{2})
	\; \mathrm{d}\lambda_{h}. \label{eq:predictprobalt}
	\end{align}
	Note that the first term in the integrand on the right hand side of (\ref{eq:predictprobalt}) does not depend on stratum or TA effects because, under the model given by (\ref{eq:Bernoulli}) -(\ref{eq:tamodel}), conditional on the household effects, under-coverage probability does not depend on stratum or TA. The second term in the integrand does not depend on TA because the model implies household effects are independent of TA effects, conditional on stratum effects. That is, TA influences the household effects only through the stratum effects. Similar arguments justify (10) in the main text.
	
	From (\ref{eq:althh2}),
	\begin{equation}
	\lambda_{h} = \alpha_{h}^{\mathrm{hh}} - (\mu + \boldX_{h}^{\mathrm{hh}'} \boldbeta^{\mathrm{hh}} + \boldX_{\mathrm{psu}[h]}^{\mathrm{psu}'}\boldbeta_{\mathrm{psu}} ). \label{eq:lambda} 
	\end{equation}
	Moreover, due to the symmetry of the normal density in the argument and the mean parameter
	\begin{align}
	\mathcal{N}(\lambda_{h}|\alpha_{s}^{\mathrm{strat}},\sigma_{\mathrm{hh}}^{2}+\sigma_{\mathrm{psu}}^{2})
	& = \mathcal{N} \left (\alpha_{h}^{\mathrm{hh}} - (\mu + \boldX_{h}^{\mathrm{hh}'}\boldbeta^{\mathrm{hh}} + \boldX_{\mathrm{psu}[h]}^{\mathrm{psu}'} \boldbeta^{\mathrm{psu}} ) | \alpha_{s}^{\mathrm{strat}},\sigma_{\mathrm{hh}}^{2}+\sigma_{\mathrm{psu}}^{2} \right ) \nonumber \\
	& = \mathcal{N} \left (\alpha_{h}^{\mathrm{hh}} |\mu + \boldX_{s}^{\mathrm{hh}'} \boldbeta^{\mathrm{hh}} + \boldX_{\mathrm{psu}[h]}^{\mathrm{psu}'}\boldbeta^{\mathrm{psu}} + \alpha_{s}^{\mathrm{strat}} , \sigma_{\mathrm{hh}}^{2}+\sigma_{\mathrm{psu}}^{2} \right). \label{eq:normaldensity}
	\end{align}
	
	Consequently, from (\ref{eq:lambda}) and (\ref{eq:normaldensity}), (\ref{eq:predictprobalt}) can be written as
	\begin{align}
	\Pr(Y=1 | & \boldXind=\boldxind,  \boldX^{\mathrm{hh}}=\boldx^{\mathrm{hh}}, \boldX^{\mathrm{psu}}=\boldx^{\mathrm{psu}}, \mathrm{TA}=t,\mathrm{strat}=s,\boldxi) = \nonumber \\
	& \int \left ( \mathrm{expit}(\alpha_{h}^{\mathrm{hh}} + \boldx^{\mathrm{xind'}}\boldbeta) \right ) %
	\mathcal{N}(\alpha_{h}^{\mathrm{hh}} | \mu + \boldx_{h}^{\mathrm{hh}'} \boldbeta^{\mathrm{hh}} + \boldx_{\mathrm{psu}[h]}^{\mathrm{psu}'} \boldbeta^{\mathrm{psu}} + \alpha_{s}^{\mathrm{strat}},\sigma_{\mathrm{hh}}^{2} + \sigma_{\mathrm{psu}}^{2}) \; \mathrm{d}\alpha_{h}^{\mathrm{hh}},
	\nonumber
	\end{align}
	where the straightforward change of variable from $\lambda_{h}$ to $\alpha_{h}^{\mathrm{hh}}$ follows from the simple relationship between them, (\ref{eq:lambda}). 
	Therefore,
	\begin{align}
	\Pr( & Y=1 | \boldXind=\boldxind,  \boldX^{\mathrm{hh}}=\boldx^{\mathrm{hh}}, \boldX^{\mathrm{psu}}=\boldx^{\mathrm{psu}}, \mathrm{TA}=t,\boldtheta) = \nonumber \\
	& \sum_{s} \Bigg (\int \left ( \mathrm{expit}(\alpha_{h}^{\mathrm{hh}} + \boldx^{\mathrm{ind}'}\boldbeta) \right ) %
	\mathcal{N}(\alpha_{h}^{\mathrm{hh}} | \mu + \boldx_{h}^{\mathrm{hh}'} \boldbeta^{\mathrm{hh}} + \boldx_{\mathrm{psu}[h]}^{\mathrm{psu}'} \boldbeta^{\mathrm{psu}} + \alpha_{s}^{\mathrm{strat}},\sigma_{\mathrm{hh}}^{2} + \sigma_{\mathrm{psu}}^{2}) \; \mathrm{d}\alpha_{h}^{\mathrm{hh}} \nonumber \\
	\times & \Pr(\mathrm{strat=s} | \boldXind=\boldxind,  \boldX^{\mathrm{hh}}=\boldx^{\mathrm{hh}}, \boldX^{\mathrm{psu}}=\boldx^{\mathrm{psu}}, \mathrm{TA}=t) \Bigg )
	\nonumber
	\end{align}
	in agreement with (10). That is, the predicted under-coverage probabilities under the model specifications given by (2-7) and (\ref{eq:Bernoulli}) to (\ref{eq:tamodel}) are identical. This result is not restricted to normally distributed random effects but holds also for other densities symmetric in the argument and the mean, such as the t-distribution. 
	
\newpage

\section{Ignorability assumptions
		\label{sec:appignore} }
	In this appendix, we formalise the ignorability assumptions which justify our modelling approach to the PES data, which includes design variables in the model but does not make use of survey weights as in the design-based approach to the analysis of complex survey data. Our approach follows the framework outlined in \citet[chapter 8]{BDA3}, but, since specific details of the PES design differ from the general treatment given by \citet{BDA3}, it seems useful to explicitly formalise our assumptions. Two features of the PES design that differ from the \citet{BDA3} set up are that the covariates are observed only for the survey sample, rather than assumed known for the full population and the number of sampling units in the population, i.e. households and individuals, is not known. Although households were selected from a household sampling frame this frame does not necessarily give an accurate count of the number of households by PSU at the time of PES fieldwork.
	
	\subsection{Notation
		\label{subsec:appnotation} }
		Although we have endeavoured to maintain consistency in notation between this appendix and the main text, we have made some slight notational changes in the interests of clarity for the current exposition. These are noted below. 
	We consider a population of size $N.$
	We let $N^{\mathrm{strat}},$ $N_{s}^{\mathrm{psu}},$ $N_{sa}^{\mathrm{hh}}$ and $N_{sah}^{\mathrm{ind}}$ denote, respectively, the number of strata, the number of primary sampling units within stratum $s,$ for $s=1,\ldots, N^{\mathrm{strat}},$ the number of households in PSU $a$ in stratum $s,$ for $a \in \{1,\ldots, N_{s}^{\mathrm{psu}} \}, s \in \{ 1,\ldots, N^{\mathrm{strata}} \} $ and the number of individuals within household $h$ in PSU $a,$ in stratum $s.$ This notation for the number of households in a PSU and the number of individuals in a household differs slightly from that used in the main text by explicitly reflecting the hierarchical structure of strata, PSU and household in the subscript notation. As noted above, the within - PSU household counts, $N_{sa}^{\mathrm{hh}}$ and the within household person count $N_{sah}^{\mathrm{ind}}$ cannot be assumed known in this application. The PES sample was  obtained by an area-based sampling scheme, as described in section 2. Note that although territorial authorities (TAs) were included in the model, they were not part of the survey design, and for the current discussion can be regarded as part of the auxiliary information included in the model. 
	
	In order to formalise the ignorability assumptions we need to consider the joint distribution of inclusion in the observed sample dataset and the survey variables.
	Consequently, we first establish notation to describe the pattern of inclusion in the sample. Note that inclusion in the observed sample dataset requires inclusion on the sampling frame, selection into the sample and response and is therefore not the same concept as selection into the sample.
	
	Reflecting the design of the PES sample, we introduce inclusion indicators for each stage of inclusion, within strata. Accordingly, let
	\begin{equation}
	\mathbf{I}_{s}^{\mathrm{psu}} = (I_{s1}^{\mathrm{psu}},\ldots, I_{sN_{s}^{\mathrm{psu}}}^{\mathrm{psu}})' \nonumber 
	\end{equation}
	be a vector of indicators for inclusion or otherwise in the sample of each of the $N_{s}^{\mathrm{psu}}$ PSUs in the stratum $s.$ Similarly, let
	\begin{equation}
	\mathbf{I}_{sa}^{\mathrm{hh}} = (I_{sa1}^{\mathrm{hh}},\ldots, I_{saN_{sa}^{\mathrm{hh}}}^{\mathrm{hh}})' \nonumber 
	\end{equation} 
	denote a vector of household inclusion indicators for the $a^{th}$ PSU in stratum $s$ and
	\begin{equation}
	\mathbf{I}_{sah}^{\mathrm{ind}} = (I_{sah1}^{\mathrm{ind}},\ldots, I_{sahN_{sah}^{\mathrm{ind}}}^{\mathrm{ind}})' \nonumber 
	\end{equation} 
	a vector of sample inclusion indicators, for individuals in the $h^{th}$ household in the PSU $a$ within stratum $s.$
	
	There are logical dependencies between the sets of inclusion indicators:
	\begin{align}
	I_{sa}^{\mathrm{psu}}=0 & \Rightarrow \boldI_{sa}^{\mathrm{hh}} = \bm{0},   \nonumber \\ %
	I_{sah}^{\mathrm{hh}} = 0 & \Rightarrow \boldI_{sah}^{\mathrm{ind}} = \bm{0}, \nonumber
	\end{align}
	where bold-faced $\bm{0}$ denotes a vector with each element equal to zero.
	
	It is also useful to define full PSU, household and individual inclusion vectors
	as
	\begin{align}
	\boldI^{\mathrm{psu}} & = (\boldI_{1}^{\mathrm{psu}}, \ldots, \boldI_{N^{\mathrm{strata}}}^{\mathrm{psu}} ),'\nonumber \\
	\boldI^{\mathrm{hh}} & = \{\boldI_{sa}^{\mathrm{hh}}, a=1,\ldots N_{s}^{\mathrm{psu}}, s=1,\ldots, N^{\mathrm{strata}}
	\}, \nonumber \\
	\boldI^{\mathrm{ind}} & = \{\boldI_{sah}^{\mathrm{ind}}, s=1,\ldots, h=1,\ldots,N_{sa}^{\mathrm{hh}},
	a=1,\ldots,N_{s}^{\mathrm{psu}},
	s= 1,\ldots,N^{\mathrm{strata}} \}.\nonumber
	\end{align}
	$\boldI^{\mathrm{hh}}$ and $\boldI^{\mathrm{ind}}$ are both column vectors, structured by PSU, and PSU and household respectively. We have not written them explicitly in that form to avoid cumbersome notation.
	
	Similarly we let
	\begin{align}
	\boldN^{\mathrm{psu}} & = (N_{1}^{\mathrm{psu}},\ldots,N_{N^{\mathrm{strata}}}^{\mathrm{psu}}),' \nonumber \\
	\boldN^{\mathrm{hh}} & = \{ N_{sa}^{\mathrm{hh}}, a= 1,\ldots, N_{s}^{\mathrm{psu}},s=1,\dots, N^{\mathrm{strata}} \}, \nonumber \\
	\boldN^{\mathrm{ind}} & = \{ N_{sah}^{\mathrm{ind}}, h=1, \ldots, N_{sa}^{\mathrm{hh}}, 
	a= 1,\ldots, N_{s}^{\mathrm{psu}},s=1,\dots, N^{\mathrm{strata}} \}, \nonumber
	\end{align}
	where $\boldN^{\mathrm{hh}}$ is a vector of household counts for the whole country, structured by PSU and strata. $\boldN^{\mathrm{ind}}$ is a vector of counts of individuals, structured by household, PSU and stratum. Neither $\boldN^{\mathrm{hh}}$ nor $\boldN^{\mathrm{ind}}$ are fully observed.
	
	Selection of PSUs is typically from a list and under the direct control of the sampler. Therefore, in typical area samples $\mathbf{I}_{s} ^{\mathrm{psu}},s=1,\ldots,N^{\mathrm{strata}}$ is directly observed and this was the case in PES. The household inclusion indicator is observed and equal to one for all selected households that respond to the survey and is observed and equal to zero for selected households that do not respond. However, in area-based sampling there is a risk that some dwellings may not be represented on the sampling frame. Consequently, there are an unknown number of households that, though part of the target population of households, were erroneously given an effective selection probability of zero. The inclusion indicator for such households is clearly zero but the number of such households is unknown. Similarly, although households in non-selected PSUs clearly have an inclusion indicator equal to zero, the total number of such households will be unknown unless the sampling frame for households is perfect in those PSUs. A useful alternative representation of the household inclusion vector is $\mathbf{I}^{\mathrm{hh}} \stackrel{\mathrm{df}}{=} (\tilde{\mathbf{I}}^{\mathrm{hh}},\mathbf{N^{\mathrm{hh}}}),$ where $\tilde{\mathbf{I}}^{\mathrm{hh}}$ represents the vector of inclusion indicators for all selected households, and all known households that were not selected, by PSU and stratum. Given $\mathbf{N}^{\mathrm{hh}}$ we could add the correct number of zero entries by PSU and stratum, to obtain the complete household inclusion vector containing an entry for every household in the target population. 
	Similarly, an alternative representation of the vector of individual inclusion indicators is $\mathbf{I}^{\mathrm{ind}} \stackrel{\mathrm{df}}{=}(\tilde{\mathbf{I}}^{\mathrm{ind}},\mathbf{N}^{\mathrm{ind}}),$ where $\tilde{\mathbf{I}}^{\mathrm{ind}}$ is the vector of recorded inclusion indicators for all respondents, and for known non-respondents from responding households, structured by household, PSU and stratum. Knowledge of $\mathbf{N}^{\mathrm{ind}}$ would allow us to construct a completed individual level inclusion vector for the target population by appending the correct number of zero entries, by household, PSU and stratum, to $\tilde{\mathbf{I}}^{\mathrm{ind}}. $ Of course, if we had complete knowledge of $\mathbf{N}^{\mathrm{ind}}$ we would have little need for a census post enumeration survey to assess the population coverage of a census. The fact that $\mathbf{N}^{\mathrm{hh}}$ and $\mathbf{N}^{\mathrm{ind}}$ are only partially observed means that the household and individual inclusion vectors are only partially observed, in contrast to the set-up assumed by \citet[chapter 8]{BDA3}.
	
	\subsection{Population model: Model for the joint distribution of outcome and covariates
		\label{subsec:apppopmodel} }
	Let $\mathbf{Y}^{\mathrm{tot}} = (Y_{1},\dots,Y_{N})'$ denote a vector of length $N$ containing census under-coverage indicators for the full target  population. Thus $Y_{i}=1$ indicates that the $i^{th}$ individual in the target population was included in the census, $Y_{i}=0,$ otherwise. In terms of the notation established in \ref{subsec:appnotation} the index $i$ represents the results of a mapping from the  stratum, PSU, household and within-household individual indices to the positive integers. We also let $\boldX_{i}$ denote a $q \times 1$ vector of covariate values for the $i^{th}$ individual in the population and let $\mathbf{X}^{\mathrm{tot}}= (\boldX_{1},\ldots,\boldX_{N})^{'}$ be an $N \times q$ matrix of covariate values for the population. In our analysis $\boldX$ includes the variables, sex, age, ethnic group, and binary indicators for \Maori descent and country of birth (denoted by $\boldXind$ in the main text). In this appendix we also consider the identity of individuals' households and household level covariates aggregated from individual covariate values as part of  $\boldX.$ That is, the definition of $\boldX$ used in this appendix, includes $\boldXind,\boldX^{\mathrm{hh}},$ as defined in the main text, as well as indicators for the household to which an individual belongs.  $\boldX$ is observed through the interview process for PES respondents but is unobserved for non-respondents. We let $\mathbf{Z}^{\mathrm{tot}}$ denote all additional available data relevant to the analysis. This includes the relationships between strata, TAs, PSUs, and households as well as covariates defined at each of these levels, except for covariates that are aggregated from the individual level covariates. Thus, in terms of the model given by (1-6) we regard $\boldX^{\mathrm{psu}}$, $\boldX^{\mathrm{ta}}$ and the stratum to TA occurrence matrix $\mathbf{W}$ as part of $\boldZ^{\mathrm{tot}}$. We regard $\mathbf{Z}^{\mathrm{tot}}$ as, in principle, known before commencing fieldwork, and is therefore not part of the data collected by PES but rather prior or auxiliary information relevant to the analysis.
	
	Our primary focus is on the relationship of census under-coverage, $Y,$ to the covariates, $\boldX,$ and TA which enters the estimation through the relationship of household to TA (via PSU and stratum). Given model parameters $\boldsymbol{\eta}=(\bm{\theta},\bm{\gamma})$ we model the joint distribution of under-coverage and covariates, conditional on the auxiliary information as 
	\begin{equation*}
	p(\boldY^{\mathrm{tot}},\boldX^{\mathrm{tot}}|\mathbf{Z}^{\mathrm{tot}},\boldsymbol{\eta})=p(\boldY^{\mathrm{tot}}|\boldX^{\mathrm{tot}},\mathbf{Z}^{\mathrm{tot}},\boldsymbol{\theta})p(\boldX^{\mathrm{tot}}|\boldZ^{\mathrm{tot}},\boldsymbol{\gamma}), \label{eq:populationmodel}
	\end{equation*}
	and assume \emph{a priori} independence for parameters of the covariate distribution and 
	the conditional distribution of the census under-coverage given the covariates, that is $	p(\bm{\eta})=p(\boldtheta)p(\boldgamma).$

	Adopting the hierarchical structure of the under-coverage model described in section 3.3 in the main text, we can write the joint conditional distribution of the under-coverage indicators as 
	\begin{equation}
	p(\boldY^{\mathrm{tot}}|\boldX^{\mathrm{tot}},\boldZ^{\mathrm{tot}},\boldtheta) =
	\prod_{s}\prod_{a}\prod_{h}\prod_{j}p(Y_{sahj}|\boldX_{sahj},\boldZ^{\mathrm{tot}},\boldtheta ).
	\label{eq:condindep}
	\end{equation}
	where $Y_{sahj}$ and $\boldX_{sahj}$ refer to coverage indicator and covariates for the $j^{th}$ person in the $h^{th}$ household within PSU $a$ in stratum $s.$
	
	The conditional independence assumptions embodied by (\ref{eq:condindep}) are critically dependent on the content of the covariate and parameter vectors. For example, if $\boldtheta$ does not include household-level intercept terms as in (3), or (\ref{eq:logitmodel}) and household is associated with tendency to respond to Census then conditional independence over individuals within households will not hold. In this case the final product term in (\ref{eq:condindep}) would need to be replaced by a model for the vector of within-household responses that explicitly modelled the associations between response for individuals in the same household.
	The conditional independence structure of (\ref{eq:condindep}) is, in fact, stronger than required for the development below but is implied by the model of (2-7), or, equivalently, (\ref{eq:Bernoulli}-\ref{eq:tamodel}) and is assumed henceforth.

	Our inferential focus is on the parameters of the hierarchical model relating under-coverage to covariates, and we therefore wish to compute the posterior distribution $p(\boldsymbol{\theta}|\mathbf{D}^{obs})$, where $\mathbf{D}^{obs}$ denotes the observed data. To define $\boldD^{\mathrm{obs}}$ it is helpful to partition $\mathbf{Y}^{\mathrm{tot}}$ and $\mathbf{X}^{\mathrm{tot}}$ into components corresponding to individuals included (denoted by superscript inc) and excluded (denoted by superscript exc) from the survey sample. Consequently, we write $\mathbf{Y}^{\mathrm{tot}}=(\mathbf{Y}^{\mathrm{inc}},\mathbf{Y}^{\mathrm{exc}})$ and $\mathbf{X}^{\mathrm{tot}}=(\mathbf{X}^{\mathrm{inc}},\mathbf{X}^{\mathrm{exc}}),$ and note that the observable data, including inclusion indicators, are $\mathbf{D}^{obs}=(\mathbf{I}^{\mathrm{psu}},\tilde{\mathbf{I}}^{\mathrm{hh}},\tilde{\mathbf{I}}^{\mathrm{ind}},\mathbf{Y}^{\mathrm{inc}},\mathbf{X}^{\mathrm{inc}}).$ The auxiliary data $\boldZ^{\mathrm{tot}}$ could also be regarded as part of $\boldD^{\mathrm{obs}},$ but we have kept it as a separate entity, that is part of the conditioning information for our models but not observed through the PES survey process.
	
	\subsection{Model for inclusion given outcome and covariates
		\label{sec:appjointmod} }
	We let $\boldphi$ denote the parameters of the conditional distribution of the inclusion indicators given $(\boldY^{\mathrm{tot}},\boldX^{\mathrm{tot}},\boldZ^{\mathrm{tot}}),$ $\bm{\psi} = (\bm{\phi},\boldsymbol{\theta},\boldgamma)$ and model the joint distribution of all inclusion indicators, outcomes and covariates, conditional on the auxiliary information, $\boldZ^{\mathrm{tot}},$ using the decomposition
	\begin{align*}
	p(\mathbf{I}^{\mathrm{psu}}, & \mathbf{I}^{\mathrm{hh}},\mathbf{I}^{\mathrm{ind}},\mathbf{Y}^{\mathrm{tot}},\mathbf{X}^{\mathrm{tot}}|\mathbf{Z}^{\mathrm{tot}},\bm{\eta})= \nonumber \\
	& p(\mathbf{I}^{\mathrm{psu}},\mathbf{I}^{\mathrm{hh}},\mathbf{I}^{\mathrm{ind}}|\mathbf{Y}^{\mathrm{tot}},\mathbf{X}^{\mathrm{tot}},
	\mathbf{Z}^{\mathrm{tot}},\boldphi)p(\mathbf{Y}^{\mathrm{tot}}|\mathbf{X}^{\mathrm{tot}},\mathbf{Z}^{\mathrm{tot}},\boldsymbol{\theta})p(\mathbf{X}^{\mathrm{tot}}|\mathbf{Z}^{\mathrm{tot}},\boldsymbol{\gamma}). \label{eq:modelbasic}
	\end{align*}
	
	In view of the relationship between PSUs, households and individuals the conditional distribution of inclusion indicators can be modelled as follows
	\begin{align*}
	p(\mathbf{I}^{\mathrm{psu}},\mathbf{I}^{\mathrm{hh}},\mathbf{I}^{\mathrm{ind}}|\mathbf{Y}^{\mathrm{tot}},\mathbf{X}^{\mathrm{tot}},
	\mathbf{Z}^{\mathrm{tot}},\boldsymbol{\phi}) = &
	p(\mathbf{I}^{\mathrm{psu}}|\mathbf{Y}^{\mathrm{tot}},\mathbf{X}^{\mathrm{tot}},\mathbf{Z}^{\mathrm{tot}},\boldsymbol{\phi}) \times \nonumber \\
	& p(\mathbf{I}^{\mathrm{hh}}|\mathbf{I}^{\mathrm{psu}},\mathbf{Y}^{\mathrm{tot}},\mathbf{X}^{\mathrm{tot}},\mathbf{Z}^{\mathrm{tot}},\boldsymbol{\phi}) \times \nonumber \\
	& p(\mathbf{I}^{\mathrm{ind}}|\mathbf{I}^{\mathrm{hh}},\mathbf{Y},\mathbf{X},\mathbf{Z},
	\boldsymbol{\phi}).
	\end{align*}
	
	We say PSU inclusion is uninformative with respect to $\mathbf{Y}^{\mathrm{tot}}$ if 
	\begin{equation}
	p(\mathbf{I}^{\mathrm{psu}}|\mathbf{Y}^{\mathrm{tot}},\mathbf{X}^{\mathrm{tot}},\mathbf{Z}^{\mathrm{tot}},%
	\boldsymbol{\phi}) = 
	p(\mathbf{I}^{\mathrm{psu}}|\mathbf{X}^{\mathrm{tot}},\mathbf{Z}^{\mathrm{tot}},\boldsymbol{\phi}). \label{eq:ignorepsu}
	\end{equation}
	Similarly, household inclusion is uninformative with respect to $\mathbf{Y}^{\mathrm{tot}}$ if 
	\begin{equation}
	p(\mathbf{I}^{\mathrm{hh}}|\mathbf{I}^{\mathrm{psu}},\mathbf{Y}^{\mathrm{tot}},\mathbf{X}^{\mathrm{tot}},\mathbf{Z}^{\mathrm{tot}},\boldsymbol{\phi}) =
	p(\mathbf{I}^{\mathrm{hh}}|\mathbf{I}^{\mathrm{psu}},\mathbf{X}^{\mathrm{tot}},\mathbf{Z}^{\mathrm{tot}},\boldsymbol{\phi})
	\label{eq:ignoredwell}
	\end{equation}
	and inclusion of individuals is ignorable with respect to $\mathbf{Y}^{\mathrm{tot}}$ if 
	\begin{align}
	p(\mathbf{I}^{\mathrm{ind}}|\mathbf{I}^{\mathrm{hh}},\mathbf{Y}^{\mathrm{tot}},\mathbf{X}^{\mathrm{tot}},\mathbf{Z}^{\mathrm{tot}},\boldsymbol{\phi}) = & p(\mathbf{I}^{\mathrm{ind}}|\mathbf{I}^{\mathrm{hh}},\mathbf{X}^{\mathrm{tot}},\mathbf{Z}^{\mathrm{tot}},\boldsymbol{\phi}). 
	\label{eq:ignoreindiv2}
	\end{align}
	
The uninformative inclusion assumptions (\ref{eq:ignorepsu})-(\ref{eq:ignoreindiv2}) are a version of the ``independence assumption" that is commonly invoked in dual systems approaches to population estimation \citep{sekar1949method,Brownonscoverage}. Simply put, by assuming uninformative inclusion, as defined above, we are assuming that the probability of inclusion in the PES does not depend on inclusion in the census. Equations (\ref{eq:ignorepsu})-(\ref{eq:ignoreindiv2}) state this assumption for each stage of selection and response. As with the conditional independence assumption of (\ref{eq:condindep}) the validity of the uninformative inclusion assumptions depends critically on the conditioning information contained in $(\boldX^{\mathrm{tot}}, \boldZ^{\mathrm{tot}})$. For example, since PSUs in urban strata were sampled at higher rate than PSUs in non-urban strata, if census coverage varied between urban and non-urban areas, PSU inclusion would be informative since PSU inclusion would be associated with census coverage. However, within strata, inclusion of PSUs depends only on a measure of size that is related to the proportion of Pacific people in the  PSUs, hence conditional on both stratum and the PSU size measure, it seems reasonable to assume PSU inclusion is uninformative. Therefore, stratum indicators and PSU size should be included in the model and we regard these variables as part of the auxiliary information, $\boldZ^{\mathrm{tot}}.$ In a multilevel model the natural way to include stratum indicators and PSU size is as covariates in the PSU level model.
	
	An advantage of specifying the uninformative inclusion assumptions in the sequential manner of (\ref{eq:ignorepsu})-(\ref{eq:ignoreindiv2}), corresponding to the stages of selection and response, is that it facilitates consideration of the plausibility of the assumptions. 

	\subsection{The posterior distribution of model parameters given the population and inclusion models
		\label{subsec:posterior} }
	Although our primary interest is in obtaining the posterior distribution for the parameters of the census coverage model, $p(\boldsymbol{\boldtheta}|\mathbf{D}^{obs})$, we first derive the joint posterior distribution for  parameters of the coverage, covariate and inclusion models, $\bm{\psi} = (\boldtheta,\boldgamma,\boldphi).$ Given  $p(\bm{\psi} | \boldD^{\mathrm{obs}} ),$ $p(\boldsymbol{\theta}|\mathbf{D}^{obs})$ can be obtained as $p(\boldtheta|\boldD^{\mathrm{obs}}) = \int p(\bm{\psi}|\mathbf{D}^{obs}) \, \mathrm{d}(\boldgamma,\boldphi).$
	
	The likelihood function for the model parameters is obtained by integrating the complete data likelihood over the unobserved components of the complete data. The likelihood is therefore given by 
	\begin{align*}
		p(\boldD^{\mathrm{obs}} | \bm{\psi}) = \int \int \int \int & \Bigg ( p(\mathbf{I}^{\mathrm{psu}},\mathbf{I}^{\mathrm{hh}},\mathbf{I}^{\mathrm{ind}}|\mathbf{Y}^{\mathrm{tot}},\mathbf{X}^{\mathrm{tot}},\mathbf{Z}^{\mathrm{tot}},\boldphi)p(\mathbf{Y}^{\mathrm{tot}}|\mathbf{X}^{\mathrm{tot}},\mathbf{Z}^{\mathrm{tot}},\boldsymbol{\theta}) \nonumber \\
	& \times p(\mathbf{X}^{tot}|\boldZ^{tot},\boldgamma) \, \mathrm{d}\boldY^{\mathrm{exc}}
	\mathrm{d}\boldX^{\mathrm{exc}} \Bigg ) \, \mathrm{d}\mathbf{N}^{\mathrm{ind}} \, 
	\mathrm{d}\mathbf{N}^{\mathrm{hh}}.
	\end{align*}
	
	If inclusion is uninformative at each stage of selection and response we have
	\begin{align}
	\begin{split}
	p(& \boldD^{\mathrm{obs}}  | \boldpsi) = \int \int \int \Bigg ( p(\mathbf{I}^{\mathrm{psu}},\mathbf{I}^{\mathrm{hh}},\mathbf{I}^{\mathrm{ind}}|\mathbf{X}^{tot},\mathbf{Z}^{tot},\boldphi)p(\mathbf{X}^{\mathrm{tot}}|\boldZ^{\mathrm{tot}},\boldgamma)  \\ %
		& ~ \phantom{\boldD^{\mathrm{obs}}  | \boldpsi) = \int \int \int (} \times \, \left ( \int p(\mathbf{Y}^{\mathrm{tot}}|\mathbf{X}^{\mathrm{tot}},\mathbf{Z}^{\mathrm{tot}},\boldsymbol{\theta}) \, \mathrm{d}\mathbf{Y}^{\mathrm{exc}} \right ) \Bigg )\mathrm{d}\mathbf{X}^{\mathrm{exc}}\, \mathrm{d}\mathbf{N}^{\mathrm{ind}} \, \mathrm{d}\mathbf{N}^{\mathrm{hh}} \ 		\end{split}
		\nonumber \\
		& = \int \int \int p(\mathbf{I}^{\mathrm{psu}},\mathbf{I}^{\mathrm{hh}},\mathbf{I}^{\mathrm{ind}}|\mathbf{X}^{\mathrm{tot}},\mathbf{Z}^{\mathrm{tot}},
	\boldphi)p(\mathbf{X}^{\mathrm{tot}}|\boldZ^{\mathrm{tot}},\boldsymbol{\boldgamma}) p(\mathbf{Y}^{\mathrm{inc}}|\mathbf{X}^{\mathrm{tot}},\mathbf{Z}^{\mathrm{tot}},\boldsymbol{\theta}) \, \mathrm{d}\mathbf{X}^{\mathrm{exc}}\, \mathrm{d}\mathbf{N}^{\mathrm{ind}} \, \mathrm{d}\mathbf{N}^{\mathrm{hh}}\,  \nonumber \\
		& = p(\mathbf{Y}^{\mathrm{inc}}|\mathbf{X}^{\mathrm{inc}},\mathbf{Z}^{\mathrm{tot}},\boldsymbol{\theta}) \int \int \int p(\mathbf{I}^{\mathrm{psu}},\mathbf{I}^{\mathrm{hh}},\mathbf{I}^{\mathrm{ind}}|\mathbf{X}^{\mathrm{tot}},\mathbf{Z}^{\mathrm{tot}},
	\boldphi)p(\mathbf{X}^{\mathrm{tot}}|\boldZ^{\mathrm{tot}},\boldsymbol{\boldgamma}) \, \mathrm{d}\mathbf{X}^{\mathrm{exc}}\, \mathrm{d}\mathbf{N}^{\mathrm{ind}} \, \mathrm{d}\mathbf{N}^{\mathrm{hh}}, 
	\label{eq:likeignore2}
	\end{align}
	where the last equality follows from the conditional independence assumptions of our coverage model (\ref{eq:condindep}) which implies that the coverage indicator for an individual can depend on that individual's covariates but not on the covariates of others. The latter assumption, which is weaker than conditional independence, would be sufficient to justify the last equality in (\ref{eq:likeignore2}).
	
	Thus, assuming uninformative inclusion at each stage of selection and response, in addition to the standard modelling assumption of conditional independence over individuals, leads to a likelihood function that separates into a component that involves only the parameters of the coverage model, $\boldtheta,$ $p(\boldY^{\mathrm{inc}} |\boldX^{\mathrm{inc}},\boldZ^{tot},\boldtheta),$ and a component that involves the remaining parameter blocks, but not $\boldtheta.$ Therefore, under the assumption of \emph{a priori} independence for the parameters of the coverage model, covariate distribution and inclusion model, the joint posterior for the model parameters is 
		\begin{align}
	p(\bm{\psi}|\boldD^{\mathrm{obs}}) &	\propto \Bigg ( p(\boldphi)p(\boldgamma)\int \int \int p(\mathbf{I}^{\mathrm{psu}},\mathbf{I}^{\mathrm{hh}},\mathbf{I}^{\mathrm{ind}}|\mathbf{X}^{\mathrm{tot}},\mathbf{Z}^{\mathrm{tot}},
\boldphi)p(\mathbf{X}^{\mathrm{tot}}|\boldZ^{\mathrm{tot}},\boldsymbol{\boldgamma}) \, \mathrm{d}\mathbf{X}^{\mathrm{exc}}\, \mathrm{d}\mathbf{N}^{\mathrm{ind}} \,
\mathrm{d}\mathbf{N}^{\mathrm{hh}} \Bigg ) \nonumber \\
& \times p(\boldtheta)p(\mathbf{Y}^{\mathrm{inc}}|\mathbf{X}^{\mathrm{inc}},\mathbf{Z}^{\mathrm{tot}},\boldsymbol{\theta})  .
	\label{eq:posterior2}
	\end{align}
	That is, the posterior for the coverage model parameters is 
	\begin{equation*}
	p(\boldtheta | \boldD^{obs} ) 
	\propto p(\boldtheta)p(\mathbf{Y}^{\mathrm{inc}}|\mathbf{X}^{\mathrm{inc}},\mathbf{Z}^{\mathrm{tot}},\boldsymbol{\theta}) 
	\end{equation*}
	and, since $p(\mathbf{Y}^{\mathrm{inc}}|\mathbf{X}^{\mathrm{inc}},\mathbf{Z}^{\mathrm{tot}},\boldsymbol{\theta})$ does not involve the inclusion indicators or the parameters of the inclusion or covariate models, the  inclusion model can be ignored in the analysis. Moreover, the conditional independence assumption (\ref{eq:condindep}) implies that $p(\mathbf{Y}^{\mathrm{inc}}|\mathbf{X}^{\mathrm{inc}},\mathbf{Z}^{\mathrm{tot}},\boldsymbol{\theta})$ has the same form as (\ref{eq:condindep}) but with the products restricted to PSUs, households and individuals included in the survey, that is
		\begin{equation}
	p(\boldY^{\mathrm{inc}}|\boldX^{\mathrm{inc}},\boldZ^{\mathrm{inc}},\boldtheta) =
	\prod_{s}\prod_{a:I_{sa}^{\mathrm{psu}}=1}\prod_{h:I_{sah}^{\mathrm{hh}}=1}\prod_{j:I_{sahj}^{\mathrm{ind}}=1}
	p(Y_{sahj}|\boldX_{sahj},\boldZ^{\mathrm{tot}},\boldtheta ).
	\nonumber
	\end{equation}
	
	If inclusion is informative at any of the sampling stages or there is dependence between the parameters of the inclusion and population models, then the inclusion model does need to be specified explicitly and included in the model fitting. \citet{pfeffermann2006multi} consider a case where the inclusion model depends on the random intercepts of a two level population model. This renders the inclusion of the second level units informative and \citet{pfeffermann2006multi} develop a conditional likelihood that explicitly incorporates inclusion indicators for each stage of inclusion and response. 
	
	Taken together, the assumptions of \emph{a priori} independence between the parameters of the inclusion and and population models, and uninformative inclusion defined by (\ref{eq:ignorepsu}) to (\ref{eq:ignoreindiv2}) imply that inclusion in the PES is ``strongly ignorable" \citep[p.203]{BDA3}. This is, in fact, a stronger assumption than strictly needed to justify ignoring the inclusion indicators in the analysis. A weaker set of assumptions that allowed the joint distribution of inclusion indicators to depend on the observed outcomes $\boldY^{\mathrm{inc}}$ but not the unobserved outcomes, $\boldY^{\mathrm{exc}}$ would suffice. However, the stronger form of the uninformative inclusion assumptions given in (\ref{eq:ignorepsu}) to (\ref{eq:ignoreindiv2}) is more readily interpretable and, as noted above, provides a link to the usual assumptions of the dual systems estimation literature.
	
	We note that from (\ref{eq:posterior2}) the posterior for the parameters of the covariate distribution does not follow straightforwardly from the observed covariate data, reflecting the impact of survey design and non-response on the observed covariate distribution of survey respondents.

\section{Additional information on model covariates}

\subsection{Hard-to-find variable}

Although presented in the main text as a household-level covariate, This binary variable is calculated at the meshblock level. 

First, for meshblocks that are present in the PES sample, a dwelling undercoverage coefficient is calculated using the ratio of number of dwellings not in the PES complete list over number of dwellings in the census frame. It is then converted into a binary variable using a threshold of 0.2, chosen from the shape of the variable' s distribution. A meshblock with a value of 1 for that variable means that the dwellings in this meshblock are more likely to be missed by PES (hard to find).

A logistic regression is then performed to predict this binary dwelling undercoverage variable from meshblock variables available in the whole country. After model selection, the selected variables are the following:
\begin{itemize}
\item	urban/rural indicator
\item	Proportion of addresses in the address register with an associated dwelling
\item Proportion of residential dwellings
\item Proportion of non-private dwellings
\item	Proportion of private dwellings at non-private dwellings
\item	Proportion of multi-dwelling addresses
\end{itemize}

The predictions from the logistic regression constitute the binary Hard-to-find variable.

\bibliography{bib18_resubmit2}